 \documentclass[
 reprint,
 amsmath,amssymb,
 aps,
 prb,
 ]{revtex4-2}

\usepackage{graphicx}
\usepackage{dcolumn}
\usepackage{bm}
\usepackage{xcolor}
\usepackage{ulem}

\begin{document}

\title{
Electronic state of vortices at twin boundaries in a nematic superconductor
}

\author{Keita~Goto$^{1,2}$}
\author{Hiroto~Adachi$^{1,2}$}
\author{Masanori~Ichioka$^{1,2}$}
\email{ichioka@okayama-u.ac.jp}



\affiliation{
$^1$Research Institute for Interdisciplinary Science, Okayama University, Okayama 700-8530, Japan\\
$^2$Department of Physics, Okayama University, Okayama 700-8530, Japan\\
}

\date{\today}

\begin{abstract}
Local electronic states of vortices in an $s \pm d$ wave nematic superconductor are studied both in the absence and presence of twin boundaries. The Bogoliubov-de Gennes theory for a tight-binding model is used with its nematicity represented by the anisotropy in the transfer integrals and attractive interactions between the nearest-neighbor sites. 
We evaluate $s$ and $d$ wave components of the pair potentials and the local density of states, and analyze the effects of nematicity on the vortex core structures with/without twin boundaries. 
We find that a single vortex trapped at the twin boundary is composed of a bound pair of fractional vortices accompanied by weakly-induced $s \pm id$ wave components, despite that such $s \pm id$ wave components do not appear in a zero magnetic field. 
The calculated spatial structures of the local electronic states are compared with the vortex image measured by STM experiments in an iron-based superconductor FeSe. 
 \end{abstract}


\maketitle


\section{Introduction}
\label{sec:introduction}

Some iron-based superconducting materials exhibit a nematic phase, an exotic twofold symmetric electronic state that emerges from the transition from a fourfold symmetric phase, and the superconductivity at low temperature shows interesting properties reflecting the nematic phase~\cite{doi:10.1073/pnas.0807325105,Bohmer_2018,doi:10.7566/JPSJ.89.102002,sym12091402}. 
Among them, a layered iron-chalcogenide FeSe becomes the nematic electronic state by the tetragonal-to-orthorhombic structural phase transition at $T_{\rm s}\sim 90 \ {\rm K}$, and shows superconductivity below $T_{\rm sc}\sim 9 \ {\rm K}$.  
In the nematic phase of FeSe, while the anisotropy of the lattice constant in the $ab$ plane is a tiny distortion of about 0.2\%~\cite{PhysRevB.87.180505}, the nematic ordering of the $d$-orbital electrons induces a large two-fold symmetric anisotropy of the Fermi surface and the superconducting gap in the nematic electronic state~\cite{PhysRevLett.113.237001,PhysRevB.90.121111,PhysRevX.8.031033,doi:10.1126/science.aal1575}, as observed in momentum space by angle-resolved photoemission spectroscopy (ARPES)~\cite{PhysRevX.8.031033} and quasiparticle interference imaging (QPI) of the scanning tunneling microscopy (STM)~\cite{doi:10.1126/science.aal1575}. 
There, two line-node-like minima of the superconducting gap are located in the long-axis directions on the nematic Fermi surface of the $\alpha$ band around the $\Gamma$ point. 

In the nematic electronic state, bulk FeSe samples exhibit a multi-domain structure consisting of two types of domains with nematic orientation along the $a$ or $b$ axis. 
Between these domains, twin boundaries appear parallel to the $(110)$ or $(-110)$ directions since the twin boundary is a crystallographic mirror plane of two neighboring domains. 
The nematic anisotropy and the twin boundary have been experimentally studied by real-space observations of electronic local density of states (LDOS) using 
STM~\cite{doi:10.1126/science.1202226, PhysRevLett.109.137004, PhysRevLett.122.077001, NatComm.9-282, doi:10.1021/acs.nanolett.3c00125, PhysRevB.99.144514, PhysRevX.5.031022,  PhysRevB.53.2835, npjQM.3-12}. 
In each domain, when magnetic fields penetrate superconductors as a vortex with a flux quantum, the vortex shows a twofold symmetric vortex core image reflecting the nematic orientation, and some vortices are trapped at the twin boundary. 

Theoretically, the nematic twofold symmetric superconducting gap structure observed in FeSe ~\cite{PhysRevX.8.031033,doi:10.1126/science.aal1575} can be described as $s \pm d$ wave pairing~\cite{doi:10.7566/JPSJ.89.102002,PhysRevX.5.031022} 
given by 
$
\Delta({\bm r},{\bm k})=\Delta_s({\bm r})+\Delta_d({\bm r})\phi_d({\bm k})
$
with $\phi_d({\bm k})=\hat{k}_x^2 -\hat{k}_y^2$, 
assuming   $\Delta_d \sim \pm \Delta_s$ in the uniform $s \pm d$ wave pairing state. 
There, the gap minimum is located at the direction of $k_x=0$ ($k_y=0$) in the momentum space for $s+d$ ($s-d$) wave pairing. 
The spatial structures of superconductivity in the vortex and the twin-boundary were phenomenologically studied by the Ginzburg-Landau (GL) theory~\cite{PhysRevB.99.144514, npjQM.3-12,PhysRevX.5.031022, PhysRevB.53.2835,doi:10.7566/JPSJ.94.023702,PhysRevB.109.094513}. 
Among them, in the $s \pm d$ wave two-component GL theory~\cite{PhysRevB.99.144514, npjQM.3-12,PhysRevX.5.031022, PhysRevB.53.2835,doi:10.7566/JPSJ.94.023702} the change of the nematic orientation at the twin boundary is considered as that ${\rm Re}\Delta_d({\bm r})$ changes the sign from negative ($s-d$ wave) to positive ($s+d$ wave) across the twin boundary, while ${\rm Re}\Delta_s({\bm r})$ remains almost constant.  
Further, it is proposed that non-zero ${\rm Im}\Delta_d({\bm r})$ appears near the twin boundary, so that the amplitude $|\Delta_d({\bm r})|$ does not vanish there, indicating local time reversal symmetry broken by the $s \pm {\it i}d$ wave superconductivity. 
This structure may lead to the emergence of fractional vortex arrays when vortices are trapped at the nematic twin boundary~\cite{PhysRevB.53.2835,10.1143/PTP.102.965,doi:10.7566/JPSJ.94.023702}. 

To confirm the exotic structures of the local time reversal symmetry broken and the fractional vortices at the nematic twin boundary suggested by the phenomenological GL theory, it is desirable to analyze them by other theoretical approaches, including estimation of electronic states to enable direct comparison with STM observations. 
As approaches in the atomic length scale, there have been previous studies on the spatial structure of twin boundaries or vortices using the tight-binding model~\cite{PhysRevB.111.054513,PhysRevB.56.R5751,PhysRevB.85.104510}. 
The twofold symmetric vortex core shape was also studied by Eilenberger theory, considering an anisotropic Fermi surface and superconducting gap~\cite{27tp-xzjb,sym12010175}.  
Following these studies, the purpose of this article is to analyze the spatial structure of superconductivity in the vortex state near the nematic twin boundary by calculations of the Bogoliubov-de Gennes equation in a tight-binding model ~\cite{PhysRevB.52.R3876,doi:10.1143/JPSJ.73.450}. 
In this study, a simple tight-binding model with nematic anisotropy in the transfer integral and attractive interactions between nearest-neighbor sites is employed to have two line-node-like gap minima on the nematic Fermi surface. 
We examine how the spatial structures of the twin boundaries and vortices appear in this simple model, estimating local electronic states in addition to the $s$ and $d$ wave components of the pair potential. 
We will see that when a vortex is trapped at the twin boundary between $s+d$ and $s-d$ wave pairing domains, a pair of fractional vortices appears by a contribution of small induced $s \pm id$ wave components, despite that such $s \pm id$ wave components do not appear in a zero magnetic field.

This paper is organized as follows. 
After the introduction in Sec. \ref{sec:introduction}, we explain our setup of the tight-binding model and the formulation of BdG equation in Sec. \ref{sec:method}. 
In Sec. \ref{sec:results}, first, we study two cases: vortex structure without twin boundary, and twin boundary structure without vortex.
Then, we consider the vortex state trapped in a twin boundary, and estimate the structure of fractional vortices in the spatial variation of the pair potential and the local electronic states.   
The last section is devoted to a summary.

\section{Formulation by Bogoliubov-de Gennens equation}
 \label{sec:method}

Our study of the superconductivity in the nematic electronic state  starts from the BCS Hamiltonian     
\begin{eqnarray} && 
{\cal H}-\mu{\cal N} 
=\sum_{i,j} 
\left(\begin{array}{cc} a^\dagger_{j,\uparrow} & a_{j,\downarrow} \\ \end{array}\right)
\left(\begin{array}{cc} K_{ji} & \Delta_{ji} \\ 
                         \Delta^\dagger_{ji} & -K^\ast_{ji} \\ \end{array}\right) 
\left(\begin{array}{c} a_{i,\uparrow} \\ a^\dagger_{i,\downarrow} \\ \end{array} \right)
\qquad 
\label{eq:BCS-H}
\end{eqnarray}
for a tight-binding model on a square lattice of atomic site $j$ in the $xy$ plane.  
The site $j$ is located at ${\bm r}_j=c(j_x,j_y)$ with the lattice constant $c$. 
$a^\dagger_{j,\sigma}$ and $a_{j,\sigma}$ are, respectively, the creation and annihilation operators of electrons with spin $\sigma=\uparrow,\downarrow$ at a site $j$.  
The pair potential $\Delta_{ji}$ and the kinetic term $K_{ji}$ are given by   
\begin{eqnarray} &&
\Delta_{ji}=\frac{1}{2}V_{ji}\langle a_{j,\uparrow} a_{i,\downarrow} \rangle , 
\label{eq:H-D}\\ && 
K_{ji}
= -t_{ji} 
\exp \left[ -{\rm  i}\frac{\pi}{\phi_0} \int^{{\bm r}_j}_{{\bm r}_i} {\bm A}({\bm r})\cdot {\rm d}{\bm r} \right]
-\delta_{ji} 
\mu, \quad 
\label{eq:H-K}
\end{eqnarray}
with the Peiels phase of the vector potential 
${\bm A}({\bm r})=\frac{1}{2}{\bm H}\times{\bm r}$ in the symmetric gauge for a magnetic field ${\bm H}=(0,0,H)$, chemical potential $\mu$, and flux quantum $\phi_0$. 
In Eq. (\ref{eq:H-D}), we consider the spin-singlet pairing satisfying $\langle a_{j,\uparrow} a_{i,\downarrow} \rangle = -\langle a_{j,\downarrow} a_{i,\uparrow} \rangle $. 

For simplicity, we assume that the transfer integral $t_{ji}$ and the pairing interaction $V_{ji}$ are only between the nearest neighbor (NN) sites $j$ and $i=j+\hat{e}$ of the square lattice ($\hat{e}=\pm\hat{x},\pm\hat{y}$). 
In a domain of the $s-d$ wave pairing, we introduce the nematic anisotropy as 
$t_{j,j\pm\hat{x}}=t_x$, $t_{j,j\pm\hat{y}}=t_y$, $V_{j,j\pm\hat{x}}=V_x$, $V_{j,j\pm\hat{y}}=V_y$.  
For the other pairs of sites, $t_{ji}=V_{ji}=0$. 

\begin{figure}[tb]
\begin{center}
\includegraphics[width=8.6cm]{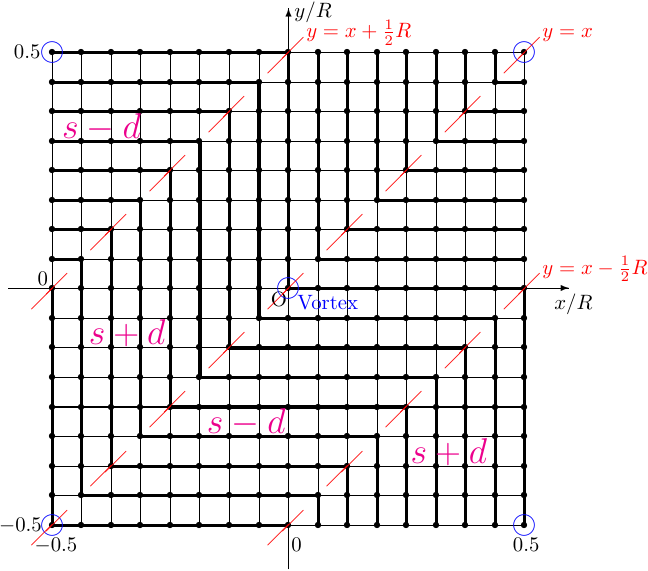}
\end{center}
\caption{\label{fig:unit-cell}
Locations of vortices and twin boundaries are schematically presented in a unit cell of square lattice with the periodic boundary condition.   
Open circles show positions of vortices at the center and corners of the unit cell. 
Diagonal dashed lines $y=x$ and $y=x \pm R/2$ present twin boundaries between $s+d$ and $s-d$ wave pairing domains. 
Width of bond lines in the square lattice indicates the nematic anisotropy in the strength of $t_{ji}$. 
}
\end{figure}

In the calculation for the vortex, we assume that two vortices are located at the center and corners of a unit cell of $N_r^2$ sites with periodic boundary condition of $\Delta_{ji}$, as shown in Fig. \ref{fig:unit-cell}.   
Then, size of the unit cell is $R_x R_y=2\phi_0/H$ with $R_x=R_y=R \equiv c N_r$. 
In the presence of twin boundaries, the location is assumed to be at diagonal lines $y=x$ and $y=x \pm R/2$ with the periodic boundary condition. 
Across the twin boundary from $s-d$ to $s+d$ wave pairing domains, nematic anisotropy changes by exchanging $t_x \leftrightarrow t_y$ and $V_x \leftrightarrow V_y$.   
We will examine how this simple tight-binding model of $t_{ji}$ and $V_{ji}$ describes electronic states of nematic superconductors with vortices and twin boundaries.  

After the Bogoliubov transformation in Eq. (\ref{eq:BCS-H}), the BdG equation is obtained as~\cite{PhysRevB.52.R3876,doi:10.1143/JPSJ.73.450}
\begin{eqnarray} 
\sum_i 
\left(\begin{array}{cc}  
  K_{ji} & \Delta_{ji} \\ 
  \Delta^\dagger_{ji}  & 
  - K^\ast_{ji} \\
\end{array} \right) 
\left( \begin{array}{c} 
  u_\alpha({\bm r}_i) \\ 
  v_\alpha({\bm r}_i) \\ 
\end{array}\right) 
= E_\alpha
 \left( \begin{array} {c}
  u_\alpha({\bm r}_j) \\ 
  v_\alpha({\bm r}_j) \\ 
\end{array}\right)  
\label{eq:BdG}
\end{eqnarray}
for the eigenstate $\alpha$. 
From the solution of the wave function  $u_\alpha({\bm r}_j)$, $v_\alpha({\bm r}_j) $, and the eigen-energy $E_\alpha$, the pair potential is calculated by the gap equation 
\begin{eqnarray}  && 
\Delta_{ji} 
=
-\frac{1}{2}V_{ji}  \sum_\alpha u_{\alpha}({\bf r}_j)  v^\ast_{\alpha}({\bf r}_i) f(-E_{\alpha})   
\label{eq:gapeq} 
\end{eqnarray}
with the Fermi distribution function $f(E)=({\rm e}^{E/T}+1)^{-1}$ at a temperature $T$. 
The calculations of Eqs. (\ref{eq:BdG}) and (\ref{eq:gapeq}) are iterated until the selfconsistent solution of $\Delta_{ji}$ is obtained. 

In our calculations, to consider the system size of $N_k^2$ unit cells with periodic boundary condition for the wave functions, we introduce magnetic Bloch states~\cite{PhysRevB.52.R3876,doi:10.1143/JPSJ.73.450}  
 \begin{eqnarray}  && 
 u_\alpha({\bm r}_j) = \tilde{u}_{\epsilon}({\bm r}_j) {\rm e}^{{\rm i}{\bm Q}_l \cdot{\bm r}_j} , \quad 
 v_\alpha({\bm r}_j) = \tilde{v}_{\epsilon}({\bm r}_j) {\rm e}^{{\rm i}{\bm Q}_l \cdot{\bm r}_j}  \quad 
\end{eqnarray}
with 
\begin{eqnarray}
 {\bm Q}_l =\frac{2 \pi}{c N_r N_k} (l_x,l_y)  .     \quad  (l_x,l_y=1,\cdots,N_k)
\end{eqnarray}
The eigenstate is labeled as $\alpha=({\bm Q}_l,\epsilon)$. $\epsilon$ is label for the eigenstate of the BdG equation for $\tilde{u}_{\epsilon}({\bm r})$ and $\tilde{v}_{\epsilon}({\bm r})$ within a unit cell of $N_r^2$ sites, under the periodic boundary condition
\begin{eqnarray} && 
\tilde{u}_\epsilon({\bm r}_j+{\bm R}) =  \tilde{u}_\epsilon({\bm r}_j) {\rm e}^{{\rm i}\chi({\bm r}_j,{\bm R})/2 } , 
\quad 
\\ && 
\tilde{v}_\epsilon({\bm r}_j+{\bm R}) =  \tilde{v}_\epsilon({\bm r}_j) {\rm e}^{-{\rm i}\chi({\bm r}_j,{\bm R})/2 }  
\end{eqnarray}
for the translation ${\bm R}= m_x {\bm R}_x +m_y {\bm R}_y$ of unit cell with ${\bm R}_x=(R_x,0)$, ${\bm R}_y=(0,R_y)$ and integers $m_x$, $m_y$. 
When there are two vortices in a unit cell at $(0,0)$ and $({\bm R}_x+{\bm R}_y)/2$, the unit vectors of the vortex lattice are given by ${\bm u}_1={\bm R}_x$ and ${\bm u}_2=({\bm R}_x+{\bm R}_y)/2$, and  
\begin{eqnarray} && 
\chi({\bm r}_j,{\bm R}) 
=-\frac{2\pi}{\phi_0}{\bm A}({\bm R}) \cdot {\bm r}_j - 2 \pi m_x(m_x-m_y)
\nonumber \\ && \qquad\qquad
+ \frac{2 \pi}{\phi_0}({\bm H}\times{\bm r}_0)\cdot{\bm R} 
\qquad 
\end{eqnarray}
in the symmetric gauge, and we set ${\bm r}_0 + \frac{1}{2}({\bm u}_1 +{\bm u}_2)=(0,0)$. 

The pair potential is decomposed to $d$-wave component $\Delta_{d}$ and extended $s$-wave one $\Delta_s$ as 
\begin{eqnarray} && 
\Delta_d({\bm r}_j)
=\Delta_{\hat{x},j}+\Delta_{-\hat{x},j}-\Delta_{\hat{y},j}-\Delta_{-\hat{y},j}
\label{eq:Delta-d}
\\ && 
\Delta_s({\bm r}_j)
=\Delta_{\hat{x},j}+\Delta_{-\hat{x},j}+\Delta_{\hat{y},j}+\Delta_{-\hat{y},j}
\label{eq:Delta-s}
\end{eqnarray}
with 
\begin{eqnarray}
\Delta_{\hat{e},j}=\Delta_{j,j+\hat{e}}\exp\left[ {\rm i} \frac{\pi}{\phi_0} \int_{{\bm r}_j}^{({\bm r}_j+{\bm r}_{j+\hat{e}})/2} {\bm A}({\bm r})\cdot{\rm d}{\bm r}\right].
\label{eq:D-bondA}
\end{eqnarray}
The phase factor in Eq. (\ref{eq:D-bondA}) is needed to satisfy the translational relation $\Delta({\bm r}_j+{\bm R})=\Delta({\bm r}_j){\rm e}^{{\rm i}\chi({\bm r}_j,{\bm R})}$.

From the selfconsistent solution of eigen states, the LDOS is calculated as~\cite{PhysRevB.52.R3876,doi:10.1143/JPSJ.73.450} 
\begin{eqnarray} && 
N(E,{\bm r}_j)=-\sum_\alpha\left\{
|u_\alpha({\bm r}_j)|^2 f'(E_\alpha -E)  
\right. \nonumber \\ && \qquad\qquad\qquad\qquad \left. 
+|v_\alpha({\bm r}_j)|^2 f'(E_\alpha +E) \right\}, 
\label{eq:LDOS}
\end{eqnarray}
where the $\delta$-function is replaced by the derivative $-f'(E)$ of $f(E)$ including thermal broadening effect in the differential tunnel conductance of STM.  

\section{Spatial structure of pair potentials and local electronic states}
 \label{sec:results}

\begin{figure}[tb]
\begin{center}
\includegraphics[width=5.0cm]{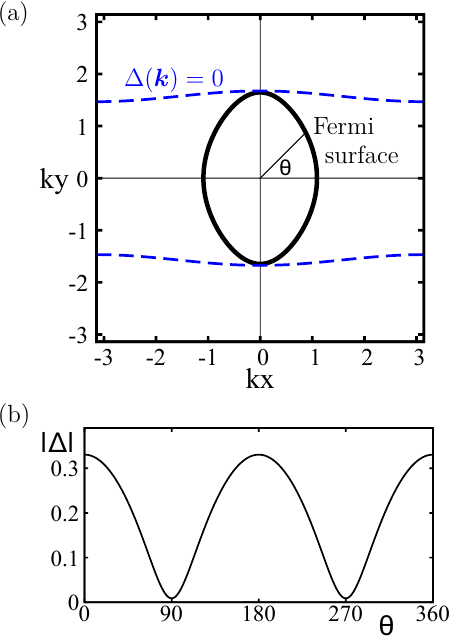}
\end{center}
\caption{\label{fig:FS}
(a)
Elliptic Fermi surface in our calculations is presented by a bold line in momentum space $(k_x,k_y)$. 
We also show dashed lines for the node positions of the superconducting gap in the uniform state, given by $\Delta({\bm k})=0$. 
(b) 
Superconducting gap $|\Delta({\bm k})|$ on the Fermi surface as a function of angle $\theta$ given in (a).
}
\end{figure}

In our calculation, we set $t_x=2$, $t_y=1$ and $\mu=-3.85$ so that the nematic Fermi surface appears around $\Gamma$ point as shown in Fig. \ref{fig:FS}(a). 
$t_y$ is used to be a unit of energy.  
To make line-node-like minima of superconducting gap at $k_x \sim 0$ on the Fermi surface, we assume that the NN site pairing interactions also have anisotropy as $V_x=-2$ and $V_y=-8$.
Equations (\ref{eq:BdG}) and (\ref{eq:gapeq}) are selfconsistently calculated at low temperature $T=0.02$ with $N_r=40$ and $N_k=3$. 

In a selfconsistent solution of the uniform state, we obtain $\Delta_s=-0.348$ and  $\Delta_d= 0.283$.
Dashed lines in Fig. \ref{fig:FS}(a) present the node position of the $s-d$ wave pair potential 
$\Delta({\bm k})
= 2\Delta_s ( \cos k_x +\cos k_y) 
 +2\Delta_d ( \cos k_x - \cos k_y)$ 
obtained from Eqs. (\ref{eq:Delta-d}) and (\ref{eq:Delta-s}).  
The superconducting gap $|\Delta({\bm k})|$ on the Fermi surface is presented in Fig. \ref{fig:FS}(b). 
There, in the long axis direction of the elliptic Fermi surface, $\Delta({\bm k}) \sim 0$. 
In the following, we calculate the spatial structure of vortices and twin boundaries in these parameters of Fig. \ref{fig:FS} as a simple model to have two line-node-like gap minima on the nematic Fermi surface.  

\subsection{Vortex without twin boundary}

\begin{figure}[tb]
\begin{center}
\includegraphics[width=8.6cm]{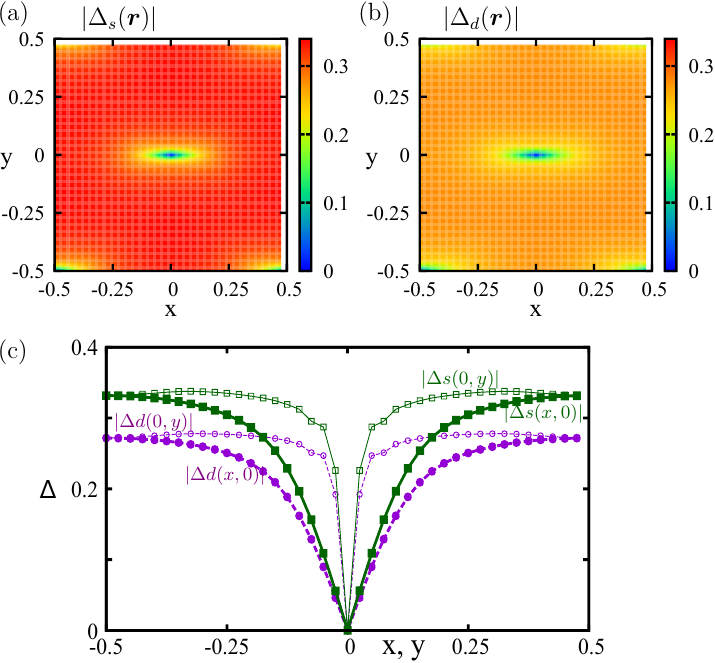}
\end{center}
\caption{\label{fig:vortexD}
Spatial structure of the vortex state is presented in real space ${\bm r}/R=(x,y)$.  
Density plots for the amplitude of the pair potential for (a) the $s$ wave component $|\Delta_s({\bm r})|$ and (b) the $d$ wave component $|\Delta_d({\bm r})|$ within a unit cell. 
(c) 
Profiles of $|\Delta_s({\bm r})|$ (solid lines) and $|\Delta_d({\bm r})|$ (dashed lines) are plotted in the path ${\bm r}/R=(x,0)$ along the $x$ axis (bold lines) and ${\bm r}/R=(0,y)$ along the $y$ axis (thin lines) from the vortex center.   
}
\end{figure}

\begin{figure}[tb]
\begin{center}
\includegraphics[width=8.6cm]{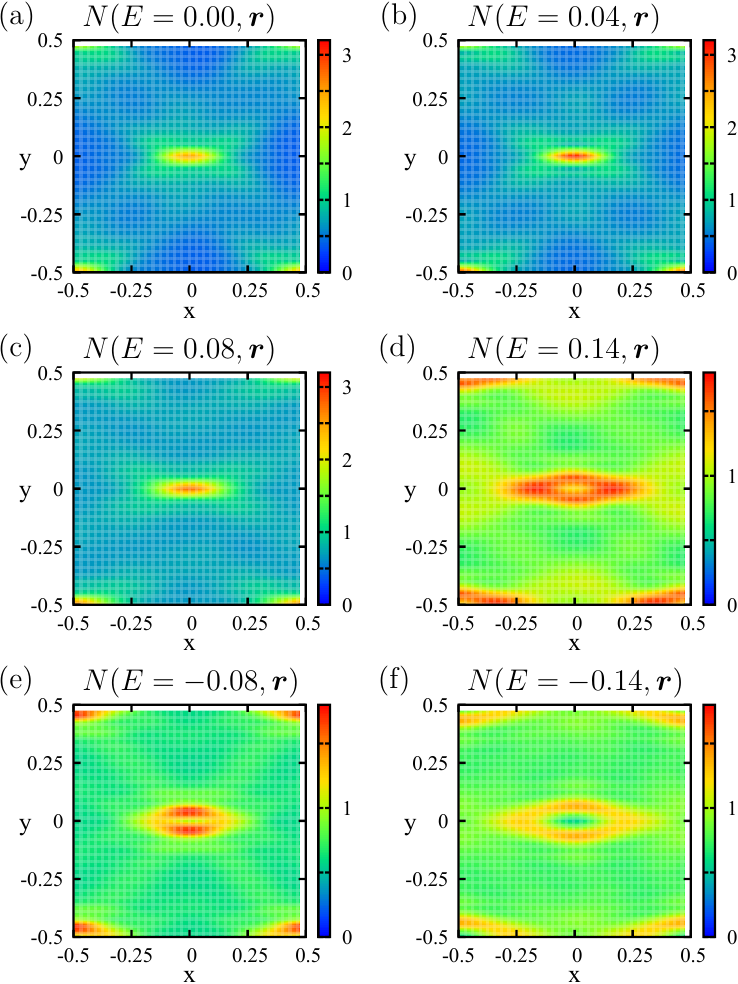}
\end{center}
\caption{\label{fig:vortexN1}
Density plots for the LDOS $N(E,{\bm r})$ at (a) $E=0$, (b) $0.04$, (c) $0.08$, (d) $0.14$, (e) $-0.08$, and (f) $-0.14$ in real space ${\bm r}/R=(x,y)$ within a unit cell.
}
\end{figure}

\begin{figure}[tb]
\begin{center}
\includegraphics[width=8.6cm]{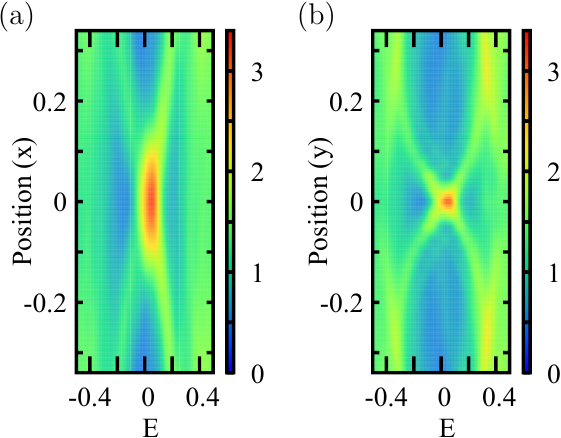}
\end{center}
\caption{\label{fig:vortexN2}
Density plots for spatial variations of the LDOS spectra $N(E,{\bf r})$ as a function of $E$ and the distance from the vortex center. Vertical axis of (a) and (b) is, respectively, along the path ${\bm r}/R=(x,0)$ and $(0,y)$. 
Vortex center is located at $(x,y)=(0,0)$. 
}
\end{figure}

First, we see the vortex structure obtained in the present tight-binding model when twin boundaries do not exist. 
The spatial structure of the pair potential is presented in Fig. \ref{fig:vortexD}. 
There, the amplitudes $|\Delta_s({\bm r})|$ is larger than  $|\Delta_d({\bm r})|$, and they show similar nematic anisotropy of the elliptic vortex core, where the core radius is larger in the $x$ direction, reflecting anisotropy of the Fermi surface shape and the superconducting gap node in Fig. \ref{fig:FS}~\cite{27tp-xzjb}. 
Around the vortex, $\Delta_d({\bm r})$ and $\Delta_s({\bm r})$ have the same phase winding $2 \pi$ with the rocked relative phase $\arg( \Delta_s^\ast \Delta_d)=\pi$ as $s-d$ wave. 

The LDOS around the vortex is presented in Fig. \ref{fig:vortexN1} to show the spatial variations within the unit cell at each $E$. 
And Fig. \ref{fig:vortexN2} shows the spectrum evolution of the LDOS along $\pm x$ or $\pm y$ directions from the vortex center.  
As seen in Figs. \ref{fig:vortexN1}(a)-(c), the low-energy bound state of the LDOS shows a two-fold symmetric distribution centered at the vortex core, which qualitatively reproduces the shape of the vortex core image observed in STM experiments~\cite{doi:10.1126/science.1202226,PhysRevLett.109.137004, PhysRevLett.122.077001,NatComm.9-282,doi:10.1021/acs.nanolett.3c00125,PhysRevB.99.144514,PhysRevX.5.031022}. 
The anisotropy of the electronic state that extends in the $x$ direction reflects the shape of the Fermi surface rather than the influence of the superconducting gap node, which was suggested in a study of Eilenberger theory~\cite{27tp-xzjb}. 

Comparing Figs. \ref{fig:vortexN1}(a)-(c), the LDOS intensity in the vortex core is highest at $E=0.04$.
This reflects the shift of the zero-energy peak to positive energy $E=0.04$ in the LDOS spectrum at the vortex center ${\bm r}=(0,0)$ in Fig. \ref{fig:vortexN2}.
This particle-hole asymmetry due to the shift of the LDOS peak at the vortex center was often seen in previous calculations of the BdG equation~\cite{PhysRevB.52.R3876,doi:10.1143/JPSJ.73.450,PhysRevLett.80.2921,Kaneko,PhysRevB.85.104510}. 
In Fig. \ref{fig:vortexN2}, discrete energy levels of the Caroli-de Gennes-Matricon state~\cite{CAROLI1964307,PhysRevLett.80.2921} are not clearly seen. 
This may be because line nodes of the superconducting gap smear the energy levels, as suggested in the $d$ wave superconductors~\cite{PhysRevB.52.R3876}.   

As $|E|$ increases further, the large LDOS region shifts from the vortex center toward the outer region~\cite{PhysRevB.43.7609,PhysRevB.53.15316,PhysRevB.59.8902,27tp-xzjb}, as seen in Figs. \ref{fig:vortexN1}(d)-(f). 
Thus, the spatial distribution of the LDOS forms a shape of two intersecting parabolas, which is explained by quasiparticle trajectories reflecting the presence of a node in the $k_y$ direction~\cite{sym12010175,27tp-xzjb}.
In FeSe, STM observations show that the large LDOS region splits into two distributions on two parallel lines at higher $|E|$~\cite{PhysRevLett.122.077001}.
At negative energies, the larger LDOS region in Figs. \ref{fig:vortexN1}(e)-(f) exhibits two similar splitting distributions, while the low intensity distribution shows remnants of two parabolas. 
In order to reproduce more similar LDOS distributions as observed by STM even at positive $E$, it is necessary to tune the tight-binding model to have the Fermi surface shape closer to that of FeSe~\cite{27tp-xzjb}.

In the spectrum of $N(E,{\bm r})$ in Fig. \ref{fig:vortexN2}, when the distance from the vortex center increases, a peak near zero energy at the vortex center splits into two peaks of positive and negative $E$, which are shifting toward the energy of the superconducting gap edge.
Along the long (short) axis of the elliptic vortex core, the splitting of the LDOS peak slowly (rapidly) occurs as a function of $x$ ($y$). 
These behaviors are qualitatively consistent with the experimental results by the STM observations~\cite{doi:10.1126/science.1202226,PhysRevLett.122.077001,PhysRevB.99.144514}.

\subsection{Twin boundary without vortex}

\begin{figure}[tb]
\begin{center}
\includegraphics[width=8.6cm]{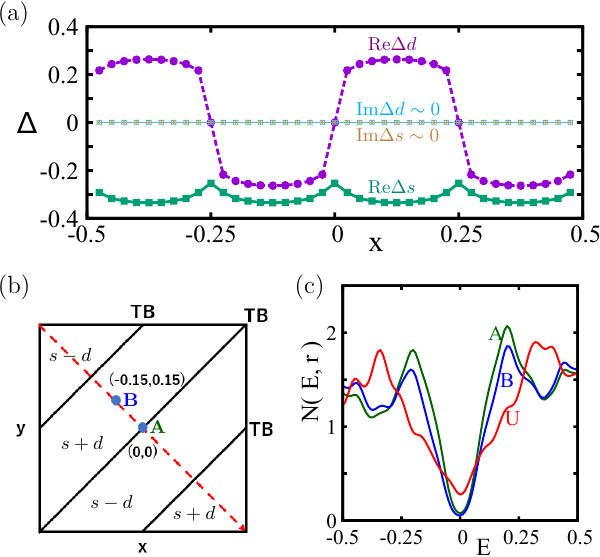}
\end{center}
\caption{\label{fig:TB}
(a) 
Spatial variation of pair potentials at a zero field $H=0$. 
We show ${\rm Re}\Delta_s({\bm r})$ and ${\rm Re}\Delta_d({\bm r})$ 
as a function of $x / R$ along a dash-dot line $y=-x$ in (b).
${\rm Im}\Delta_s({\bm r}) \sim 0$ and  ${\rm Im}\Delta_d({\bm r})\sim 0$. 
(b)
The square region of a unit cell in our calculation is schematically presented. 
The twin boundaries (TB) are located at diagonal solid lines, separating $s+d$ and $s-d$ wave pairing domains. 
Periodic boundary conditions are applied.  
(c) 
Spectra of LDOS $N(E,{\bm r})$ as a function of energy $E$. 
Line A is for just on the twin boundary ${\bm r}=(0,0)$, and line B is for a middle point ${\bm r}/R=(-0.15, 0.15)$ between two neighboring twin boundaries. 
The positions of A and B are also presented in (b). 
Line U is for $N(E)$ in the uniform state without twin boundaries. 
}
\end{figure}

Next, we study spatial structures of the twin boundaries appearing in the present model at a zero field without vortex.  
The twin boundaries and $s \pm d$ wave pairing domains are set by the anisotropies of $t_{ji}$ and $V_{ji}$ as noted in Sec. \ref{sec:method}. 
Figure \ref{fig:TB}(a) shows the spatial variation of the pair potential along the perpendicular direction to the twin boundary. 
There, the $s$ wave component ${\rm Re}\Delta_s({\bm r})$ has a negative sign, and the amplitude $|\Delta_s({\bm r})|$ shows weak suppression at the twin boundary. 
The $d$ wave component ${\rm Re}\Delta_d({\bm r})$ changes the sign from ${\rm Re}\Delta_d({\bm r}) \sim {\rm Re}\Delta_s({\bm r})$ as $s+d$ wave to ${\rm Re}\Delta_d({\bm r}) \sim -{\rm Re}\Delta_s({\bm r})$ as $s-d$ wave across the twin boundary. 
When the sign change of ${\rm Re}\Delta_d({\bm r})$ occurs through zero, the amplitude $|\Delta_d({\bm r})| \rightarrow 0$ at the twin boundary, because the imaginary part reduces to ${\rm Im}\Delta_d({\bm r}) \sim 0$ in the selfconsistent solution, even when iterations of calculations start from some initial states with finite values of ${\rm Im}\Delta_d({\bm r})$.

This is a different situation from results by $s \pm d$ wave GL theory~\cite{PhysRevB.99.144514, npjQM.3-12,PhysRevX.5.031022, PhysRevB.53.2835,doi:10.7566/JPSJ.94.023702}, where non-zero ${\rm Im}\Delta_d({\bm r})$ largely appears near the twin boundary as $s \pm id$ wave, so that the amplitude $|\Delta_d({\bm r})|$ does not vanish there. 
As a possible reason for ${\rm Im}\Delta_d({\bm r}) \sim 0$ in the present tight-binding model, the spatial variation of the superconductivity near the twin boundary is evaluated in the atomic scale, and the spacing of two neighboring twin boundaries, $c N_r / 2 \sqrt{2}$, may be too short to develop an additional component of ${\rm Im}\Delta_d({\bm r})$. 
Furthermore, while the nematic anisotropy of $t_{ji}$ and $V_{ji}$ suddenly changes through one site of the twin boundary in the present simple model, it is also worth considering a mechanism that generates an effective pairing force stabilizing the $s \pm id$ wave superconductivity in neighbor regions of twin boundaries by changing the nematic anisotropy.
Regarding the degrees of freedom of superconductivity, for simplicity, our model considers the pair potential $\Delta_{ji}$ only between NN sites, and decomposes them to the extended $s$ wave pairing component $ \cos k_x + \cos k_y$ and the $d$ wave component $\cos k_x - \cos k_y$. 
However, it may be necessary to consider also the other pairing components, such as the on-site $s$ wave component $\Delta_{jj}$ and the possible pairings in orbital degrees of freedom, to investigate possible mechanics for local stability of $s \pm id$ wave state. 
While the development of the microscopic model to induce large ${\rm Im}\Delta_d({\bm r})$ near the twin boundary belongs to future work, it is also important to consider how the local appearance of large ${\rm Im}\Delta_d({\bm r})$ changes the spatial structure of vortices at the twin boundary. 
For the purpose, we need to know the structure of the vortex states at twin boundaries when the generation of ${\rm Im}\Delta_d({\bm r})$ near the twin boundary is not only easy but also difficult. 
Thus, in the following, we study the electronic states of the twin boundary and vortex state under the present tight-binding model, where ${\rm Im}\Delta_d({\bm r})$ does not appear at twin boundaries in a zero magnetic field, as in Fig. \ref{fig:TB}(a).

To see the local electronic states, we show the spectra of the LDOS $N(E,{\bm r})$ in Fig. \ref{fig:TB}(c). 
In a uniform state without twin boundaries (line U), $N(E,{\bm r})$ shows a V-shape spectrum which linearly increases from the Fermi energy, reflecting line-node-like minima of the superconducting gap.  
Finite LDOS at $E=0$ in Fig. \ref{fig:TB}(c) is due to the thermal smearing of the factor $f'(E)$ in Eq. (\ref{eq:LDOS}). 

When twin boundaries are present, low-energy gaps appear in the spectrum at the twin boundary (line A). 
Eigen-energy $E_\alpha$ is absent within the low-energy gaps. 
This may reflect the disappearance of the $d$ wave component of the pair potential at twin boundaries, resulting in a change from the line-node-like gap minima of $s \pm d$ wave to the finite gap of $s$ wave $\Delta_s({\bm r})$. 
This $s$ wave-like low-energy gap is also found at the midpoint between two neighboring twin boundaries (line B). 
This is likely due to the short spacing of the neighboring twin boundaries in our calculations.
There, since the wave functions of electrons broadly reflect the superconducting state in neighboring regions, the spatially oscillating $\Delta_d({\bm r})$ across the twin boundary gives a weak contribution to $N(E,{\bm r})$ as the spatial average, rather than the local $s \pm d$ wave pair potential with line-node-like gap minima. 
Thus, the electrons seem to sense dominantly only $\Delta_s({\bm r})$. 
The STM spectrum suggesting the low-energy gaps was observed at the region of short spacing of the neighboring twin boundaries in FeSe~\cite{PhysRevX.5.031022}. 

 \subsection{Vortex trapped at a twin boundary}

\begin{figure}[tb]
\begin{center}
\includegraphics[width=8.6cm]{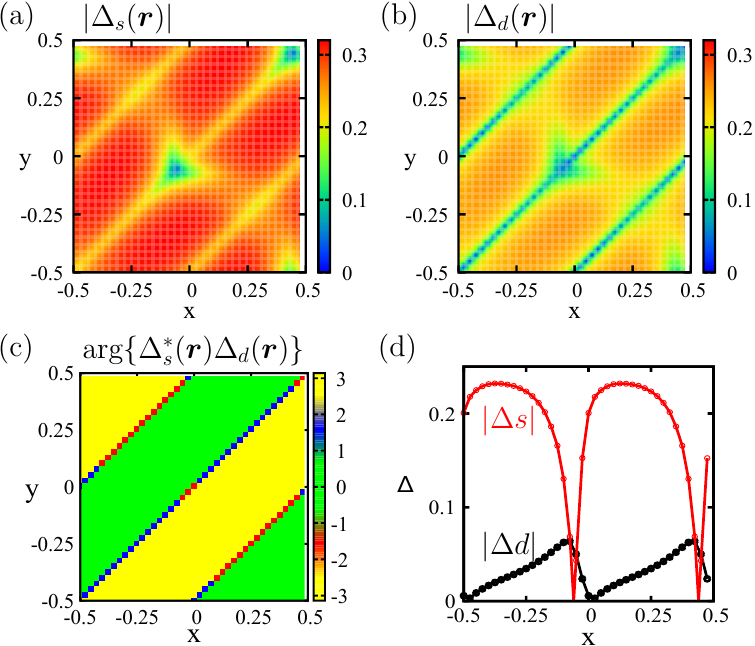}
\end{center}
\caption{\label{fig:TBvortexD}
Spatial structure of the vortex state trapped at a twin boundary is presented in real space ${\bm r}/R=(x,y)$ within a unit cell.  
Amplitude of the pair potential for (a) the $s$ wave component $|\Delta_s({\bm r})|$ and (b) the $d$ wave component $|\Delta_d({\bm r})|$. 
(c) Relative phase $ \arg \{\Delta_s^\ast({\bm r}) \Delta_d({\bm r}) \}$. 
(d) $|\Delta_s({\bm r})|$ and $|\Delta_d({\bm r})|$ are plotted along a twin boundary on a diagonal line $y=x$. 
}
\end{figure}
\begin{figure}[tb]
\begin{center}
\includegraphics[width=8.6cm]{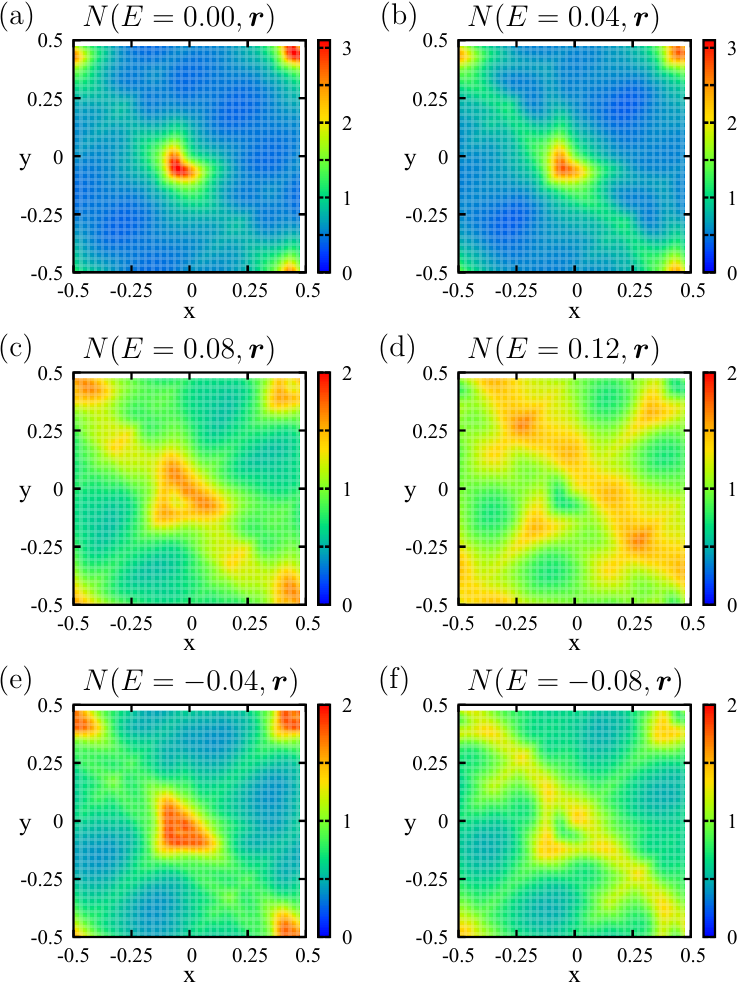}
\end{center}
\caption{\label{fig:TBvortexN1}
Density plots for the LDOS $N(E,{\bm r})$ at (a) $E=0$, (b) $0.04$, (c) $0.08$, (d) $0.12$, (e) $-0.04$, and (f) $-0.08$ in real space ${\bm r}/R=(x,y)$ within a unit cell, when vortex is trapped at a twin boundary.   
}
\end{figure}

\begin{figure}[tb]
\begin{center}
\includegraphics[width=8.6cm]{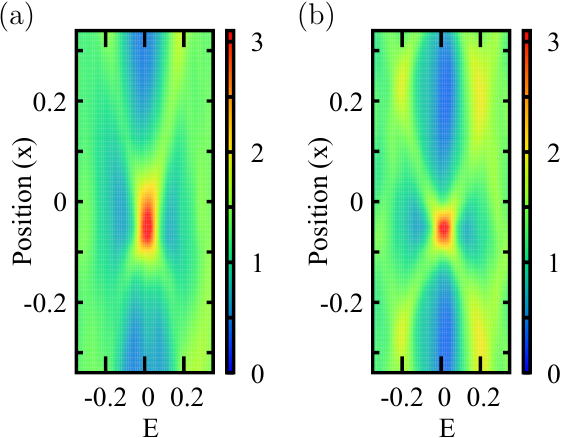}
\end{center}
\caption{\label{fig:TBvortexN2}
Density plots for spatial variations of the LDOS spectra $N(E,{\bf r})$ as a function of $E$ and the distance from the vortex center. 
Vertical axis of (a) and (b) is, respectively, the path ${\bm r}/R=(x,-0.05)$ along the sites of $x$ direction and ${\bm r}/R=(x,x)$ along a twin boundary. 
Vortex center of $\Delta_s({\bm r})$ is located at ${\bm r}/R=(-0.06,-0.06)$. 
}
\end{figure}

Lastly, we study the spatial structure of vortices trapped in a twin boundary of Fig. \ref{fig:TB}. 
Figure \ref{fig:TBvortexD} presents the spatial structure of the order  parameters. 
The $s$ wave component $|\Delta_s({\bm r})|$ in Fig. \ref{fig:TBvortexD}(a) is suppressed near the vortex center $(-0.060, -0.060)$, slightly shifted from $(0,0)$. 
The coordinate of the vortex center is estimated by the interpolation between sites.
Slight suppression is also seen at the twin boundaries on the diagonal lines.
The vortex core near the center of the figure has a shape that extends to the rightward and upward from the vortex center. 
This is consistent to the results of the GL theory coupled with nematic orders~\cite{PhysRevB.109.094513}. 
This vortex core shape reflects the fact that the lower right side of the vortex is the $s-d$ wave domain, which has the same nematic anisotropy as in Fig. \ref{fig:vortexD} and an elliptical vortex core shape elongated in the $x$ direction.
On the other hand, the upper left side of the vortex is the $s+d$ wave domain, where the nematic anisotropy is rotated $90^\circ$, resulting in an elliptical vortex core extending in the $y$ direction.

For the $d$-wave component $|\Delta_d({\bm r})|$ in Fig. \ref{fig:TBvortexD}(b), we see significant suppression at the twin boundary where the $s+d$ wave state is converted to the $s-d$ wave state. 
The vortex core shape appears to be similar to that of $|\Delta_s({\bm r})|$.
However, it is noted that the vortex center of $\Delta_d({\bm r})$ is located at $(0.016,0.016)$ slightly deviating from that of $\Delta_s({\bm r})$.
This can also be seen from the relative phase $ \arg \{\Delta_s^\ast({\bm r}) \Delta_d({\bm r}) \}$ in Fig. \ref{fig:TBvortexD}(c). 
There, a jump of the relative phase between $0$ and $\pi$ is seen at the twin boundary between the $s+d$ and $s-d$ wave domains.
Just on the twin boundary, the relative phase is $\pi/2$ or $-\pi/2$, indicating $s \pm id$ states, since small imaginary parts of $\Delta_d({\bm r})$ appear in the vortex state. 
Between (outside) the closely spaced vortices of $\Delta_s({\bm r})$ and $\Delta_d({\bm r})$, the relative phase is $- \pi /2$ ($ \pi /2$) on the twin boundary. 
Thus, the winding of the relative phase is $-2 \pi$ and $2 \pi$ around the vortex of $\Delta_s$ and $\Delta_d$, respectively. 
Therefore, paired fractional vortices of $\Delta_s({\bm r})$ and $\Delta_d({\bm r})$ appear at the twin boundary due to the induced local $s \pm id$ states.  
Figure \ref{fig:TBvortexD}(d) shows $|\Delta_s({\bm r})|$ and $|\Delta_d({\bm r})|$ along the diagonal twin boundary to show that the vortex centers of $\Delta_s({\bm r})$ and $\Delta_d({\bm r})$ are located at different positions, forming fractional quantum vortices, and also that the vortex structure is asymmetric along the twin boundary.

The spatial structure of the LDOS is shown in Fig. \ref{fig:TBvortexN1} for several cases of $E$.
The zero-energy LDOS in Fig. \ref{fig:TBvortexN1}(a) is a bound state localized in the vortex core of $\Delta_s({\bm r})$ at the vortex center ${\bm r}=(-0.060,-0.060)$. 
Reflecting the spatial structure of $|\Delta_s({\bm r})|$ in Fig. \ref{fig:TBvortexD}(a), the spatial distribution of the high intensity region in Figs. \ref{fig:TBvortexN1}(a)-(b) elongates in the $+x$ and $+y$ directions. 
The vortex core image of $N(E=0,{\bm r})$ qualitatively reproduces the shape of the local electronic state observed by STM~\cite{PhysRevLett.122.077001}.  
As $|E|$ increases, the high intensity region of the LDOS shifts from the vortex center toward the outer region, as seen in Fig. \ref{fig:TBvortexN1}(c)-(f).

Figure \ref{fig:TBvortexN2}(a) shows the LDOS spectrum $N(E,{\bm r})$ along the path ${\bm r}=(x, -0.050)$ in the $x$ direction from the site nearest to the vortex center. 
In the behavior about the splitting of the low-energy peak into two peaks of positive and negative $E$, the region at $x>-0.050$ and $x < -0.050$, respectively, corresponds to those of Fig. \ref{fig:vortexN2}(a) along the long axis and Fig. \ref{fig:vortexN2}(b) along the short axis of the elliptic vortex core.  
When the twin boundary exists, the tails of zero-energy LDOS from the vortex center along $+x$ direction in Figs. \ref{fig:TBvortexN1}(a) and \ref{fig:TBvortexN2}(a) becomes shorter compared to that in Figs. \ref{fig:vortexN1}(a) and \ref{fig:vortexN2}(a).  
The peak of spectra $N(E,{\bm r})$ at the vortex center in Fig. \ref{fig:TBvortexN2} is located at $E=0.015$, which is a smaller shift compared to the positive shift of $0.04$ in Fig. \ref{fig:vortexN2}. 
These trends may arise because the confinement of the electronic state within the vortex core is weaker when the vortex is trapped in a wide region of weaker superconductivity at the twin boundary. 

Figure \ref{fig:TBvortexN2}(b) shows spectrum $N(E,{\bm r})$ along a twin boundary $y=x$. 
There, we see that the splitting of the low-energy peak starts from the vortex center of $\Delta_s({\bm r})$ at $x=y=-0.060$. 
Zero-energy bound states are not seen at the vortex center of $\Delta_d({\bm r})$ at $x=y=0.016$. 

\section{Summary}
 \label{sec:summary}

We have theoretically studied the spatial structure of the pair potential and the local electronic state for the vortex states with and without twin boundaries in nematic superconductors, and compared them with STM observations in FeSe. 
We used a tight-binding model with nematic anisotropy in the hopping and attractive interactions between nearest-neighbor sites, tuning parameters to have line-node-like gap minima on the nematic Fermi surface. 
At the diagonal lines of the twin boundaries, it is assumed that the nematic anisotropy between the $x$ and $y$ axis directions is exchanged. 
Solving the Bogoliubov-de Gennes equation in this simple tight-binding model, we examine how the properties of the vortex states reflect contributions of nematic anisotropy and the twin boundaries. 

In the absence of the twin boundary, we study the spatial structure of the electronic LDOS $N(E,{\bm r})$ around the elliptic vortex core, which reflects the nematic Fermi surface shape, and show the energy dependence of the vortex core image in the LDOS. 
At higher energies, larger intensities of the LDOS appear on the quasiparticle trajectories reflecting line-node-like gap minima~\cite{sym12010175,27tp-xzjb}. 
We also confirmed the difference in spectral evolution around the vortex along the long and short axes of the elliptic vortex core, which are qualitatively consistent with the STM observation~\cite{doi:10.1126/science.1202226,PhysRevLett.122.077001,PhysRevB.99.144514}.

In the presence of the twin boundary, we consider a situation in which the $s \pm d$ wave pairing regions are alternating across the twin boundary. 
In our calculation under a zero magnetic field, we do not find the local $s \pm id$ wave pairing states with broken time reversal symmetry that is suggested by the two-component $s \pm d$ wave GL theory~\cite{PhysRevB.99.144514, npjQM.3-12,PhysRevX.5.031022, PhysRevB.53.2835,doi:10.7566/JPSJ.94.023702}. 
Nevertheless, in the vortex states under magnetic fields, we find that a vortex trapped at the twin boundary is composed of a bound pair of fractional vortices accompanied by a weakly-induced $s \pm id$ wave components, which is in line with the finding by the two-component $s \pm d$ wave GL theory~\cite{doi:10.7566/JPSJ.94.023702}. 
Finally, the zero-energy LDOS distributions around the vortex at the twin boundary are qualitatively consistent with the STM observation~\cite{PhysRevLett.122.077001}. 

While the above-mentioned properties of vortices and twin boundaries were obtained using a simple tight-binding model with nematic anisotropy in the hopping and attractive interactions between nearest-neighbor sites, this information will be useful for future detailed studies on the vortex states at twin boundaries, providing a new perspective for the characteristics of nematic superconductors. 
A possible extention of the present tight binding model to a more realistic model for FeSe, which is left to future studies, includes to allow for the appearance of large $s \pm id$ wave components at the twin boundaries as well as allow for the realization of fractional vortices as suggested by the two-componnent GL theory.

\begin{acknowledgments}
This work was supported by JSPS KAKENHI Grant Numbers JP21K03471 and JP26K07000.
\end{acknowledgments}



\begin{thebibliography}{34}%
\makeatletter
\providecommand \@ifxundefined [1]{%
 \@ifx{#1\undefined}
}%
\providecommand \@ifnum [1]{%
 \ifnum #1\expandafter \@firstoftwo
 \else \expandafter \@secondoftwo
 \fi
}%
\providecommand \@ifx [1]{%
 \ifx #1\expandafter \@firstoftwo
 \else \expandafter \@secondoftwo
 \fi
}%
\providecommand \natexlab [1]{#1}%
\providecommand \enquote  [1]{``#1''}%
\providecommand \bibnamefont  [1]{#1}%
\providecommand \bibfnamefont [1]{#1}%
\providecommand \citenamefont [1]{#1}%
\providecommand \href@noop [0]{\@secondoftwo}%
\providecommand \href [0]{\begingroup \@sanitize@url \@href}%
\providecommand \@href[1]{\@@startlink{#1}\@@href}%
\providecommand \@@href[1]{\endgroup#1\@@endlink}%
\providecommand \@sanitize@url [0]{\catcode `\\12\catcode `\$12\catcode
  `\&12\catcode `\#12\catcode `\^12\catcode `\_12\catcode `\%12\relax}%
\providecommand \@@startlink[1]{}%
\providecommand \@@endlink[0]{}%
\providecommand \url  [0]{\begingroup\@sanitize@url \@url }%
\providecommand \@url [1]{\endgroup\@href {#1}{\urlprefix }}%
\providecommand \urlprefix  [0]{URL }%
\providecommand \Eprint [0]{\href }%
\providecommand \doibase [0]{https://doi.org/}%
\providecommand \selectlanguage [0]{\@gobble}%
\providecommand \bibinfo  [0]{\@secondoftwo}%
\providecommand \bibfield  [0]{\@secondoftwo}%
\providecommand \translation [1]{[#1]}%
\providecommand \BibitemOpen [0]{}%
\providecommand \bibitemStop [0]{}%
\providecommand \bibitemNoStop [0]{.\EOS\space}%
\providecommand \EOS [0]{\spacefactor3000\relax}%
\providecommand \BibitemShut  [1]{\csname bibitem#1\endcsname}%
\let\auto@bib@innerbib\@empty
\bibitem [{\citenamefont {Hsu}\ \emph {et~al.}(2008)\citenamefont {Hsu},
  \citenamefont {Luo}, \citenamefont {Yeh}, \citenamefont {Chen}, \citenamefont
  {Huang}, \citenamefont {Wu}, \citenamefont {Lee}, \citenamefont {Huang},
  \citenamefont {Chu}, \citenamefont {Yan},\ and\ \citenamefont
  {Wu}}]{doi:10.1073/pnas.0807325105}%
  \BibitemOpen
  \bibfield  {author} {\bibinfo {author} {\bibfnamefont {F.-C.}\ \bibnamefont
  {Hsu}}, \bibinfo {author} {\bibfnamefont {J.-Y.}\ \bibnamefont {Luo}},
  \bibinfo {author} {\bibfnamefont {K.-W.}\ \bibnamefont {Yeh}}, \bibinfo
  {author} {\bibfnamefont {T.-K.}\ \bibnamefont {Chen}}, \bibinfo {author}
  {\bibfnamefont {T.-W.}\ \bibnamefont {Huang}}, \bibinfo {author}
  {\bibfnamefont {P.~M.}\ \bibnamefont {Wu}}, \bibinfo {author} {\bibfnamefont
  {Y.-C.}\ \bibnamefont {Lee}}, \bibinfo {author} {\bibfnamefont {Y.-L.}\
  \bibnamefont {Huang}}, \bibinfo {author} {\bibfnamefont {Y.-Y.}\ \bibnamefont
  {Chu}}, \bibinfo {author} {\bibfnamefont {D.-C.}\ \bibnamefont {Yan}},\ and\
  \bibinfo {author} {\bibfnamefont {M.-K.}\ \bibnamefont {Wu}},\ }\bibfield
  {title} {\bibinfo {title} {{Superconductivity in the PbO-type structure
  $\alpha$-FeSe}},\ }\href {https://doi.org/10.1073/pnas.0807325105} {\bibfield
   {journal} {\bibinfo  {journal} {Proceedings of the National Academy of
  Sciences}\ }\textbf {\bibinfo {volume} {105}},\ \bibinfo {pages} {14262}
  (\bibinfo {year} {2008})}\BibitemShut {NoStop}%
\bibitem [{\citenamefont {Böhmer}\ and\ \citenamefont
  {Kreisel}(2017)}]{Bohmer_2018}%
  \BibitemOpen
  \bibfield  {author} {\bibinfo {author} {\bibfnamefont {A.~E.}\ \bibnamefont
  {Böhmer}}\ and\ \bibinfo {author} {\bibfnamefont {A.}~\bibnamefont
  {Kreisel}},\ }\bibfield  {title} {\bibinfo {title} {{Nematicity, magnetism
  and superconductivity in FeSe}},\ }\href
  {https://doi.org/10.1088/1361-648X/aa9caa} {\bibfield  {journal} {\bibinfo
  {journal} {Journal of Physics: Condensed Matter}\ }\textbf {\bibinfo {volume}
  {30}},\ \bibinfo {pages} {023001} (\bibinfo {year} {2017})}\BibitemShut
  {NoStop}%
\bibitem [{\citenamefont {Shibauchi}\ \emph {et~al.}(2020)\citenamefont
  {Shibauchi}, \citenamefont {Hanaguri},\ and\ \citenamefont
  {Matsuda}}]{doi:10.7566/JPSJ.89.102002}%
  \BibitemOpen
  \bibfield  {author} {\bibinfo {author} {\bibfnamefont {T.}~\bibnamefont
  {Shibauchi}}, \bibinfo {author} {\bibfnamefont {T.}~\bibnamefont
  {Hanaguri}},\ and\ \bibinfo {author} {\bibfnamefont {Y.}~\bibnamefont
  {Matsuda}},\ }\bibfield  {title} {\bibinfo {title} {{Exotic Superconducting
  States in FeSe-based Materials}},\ }\href
  {https://doi.org/10.7566/JPSJ.89.102002} {\bibfield  {journal} {\bibinfo
  {journal} {Journal of the Physical Society of Japan}\ }\textbf {\bibinfo
  {volume} {89}},\ \bibinfo {pages} {102002} (\bibinfo {year}
  {2020})}\BibitemShut {NoStop}%
\bibitem [{\citenamefont {Kreisel}\ \emph {et~al.}(2020)\citenamefont
  {Kreisel}, \citenamefont {Hirschfeld},\ and\ \citenamefont
  {Andersen}}]{sym12091402}%
  \BibitemOpen
  \bibfield  {author} {\bibinfo {author} {\bibfnamefont {A.}~\bibnamefont
  {Kreisel}}, \bibinfo {author} {\bibfnamefont {P.~J.}\ \bibnamefont
  {Hirschfeld}},\ and\ \bibinfo {author} {\bibfnamefont {B.~M.}\ \bibnamefont
  {Andersen}},\ }\bibfield  {title} {\bibinfo {title} {{On the Remarkable
  Superconductivity of FeSe and Its Close Cousins}},\ }\href
  {https://www.mdpi.com/2073-8994/12/9/1402} {\bibfield  {journal} {\bibinfo
  {journal} {Symmetry}\ }\textbf {\bibinfo {volume} {12}},\ \bibinfo {pages}
  {1402} (\bibinfo {year} {2020})}\BibitemShut {NoStop}%
\bibitem [{\citenamefont {B\"ohmer}\ \emph {et~al.}(2013)\citenamefont
  {B\"ohmer}, \citenamefont {Hardy}, \citenamefont {Eilers}, \citenamefont
  {Ernst}, \citenamefont {Adelmann}, \citenamefont {Schweiss}, \citenamefont
  {Wolf},\ and\ \citenamefont {Meingast}}]{PhysRevB.87.180505}%
  \BibitemOpen
  \bibfield  {author} {\bibinfo {author} {\bibfnamefont {A.~E.}\ \bibnamefont
  {B\"ohmer}}, \bibinfo {author} {\bibfnamefont {F.}~\bibnamefont {Hardy}},
  \bibinfo {author} {\bibfnamefont {F.}~\bibnamefont {Eilers}}, \bibinfo
  {author} {\bibfnamefont {D.}~\bibnamefont {Ernst}}, \bibinfo {author}
  {\bibfnamefont {P.}~\bibnamefont {Adelmann}}, \bibinfo {author}
  {\bibfnamefont {P.}~\bibnamefont {Schweiss}}, \bibinfo {author}
  {\bibfnamefont {T.}~\bibnamefont {Wolf}},\ and\ \bibinfo {author}
  {\bibfnamefont {C.}~\bibnamefont {Meingast}},\ }\bibfield  {title} {\bibinfo
  {title} {{Lack of coupling between superconductivity and orthorhombic
  distortion in stoichiometric single-crystalline FeSe}},\ }\href
  {https://doi.org/10.1103/PhysRevB.87.180505} {\bibfield  {journal} {\bibinfo
  {journal} {Phys. Rev. B}\ }\textbf {\bibinfo {volume} {87}},\ \bibinfo
  {pages} {180505} (\bibinfo {year} {2013})}\BibitemShut {NoStop}%
\bibitem [{\citenamefont {Nakayama}\ \emph {et~al.}(2014)\citenamefont
  {Nakayama}, \citenamefont {Miyata}, \citenamefont {Phan}, \citenamefont
  {Sato}, \citenamefont {Tanabe}, \citenamefont {Urata}, \citenamefont
  {Tanigaki},\ and\ \citenamefont {Takahashi}}]{PhysRevLett.113.237001}%
  \BibitemOpen
  \bibfield  {author} {\bibinfo {author} {\bibfnamefont {K.}~\bibnamefont
  {Nakayama}}, \bibinfo {author} {\bibfnamefont {Y.}~\bibnamefont {Miyata}},
  \bibinfo {author} {\bibfnamefont {G.~N.}\ \bibnamefont {Phan}}, \bibinfo
  {author} {\bibfnamefont {T.}~\bibnamefont {Sato}}, \bibinfo {author}
  {\bibfnamefont {Y.}~\bibnamefont {Tanabe}}, \bibinfo {author} {\bibfnamefont
  {T.}~\bibnamefont {Urata}}, \bibinfo {author} {\bibfnamefont
  {K.}~\bibnamefont {Tanigaki}},\ and\ \bibinfo {author} {\bibfnamefont
  {T.}~\bibnamefont {Takahashi}},\ }\bibfield  {title} {\bibinfo {title}
  {{Reconstruction of Band Structure Induced by Electronic Nematicity in an
  FeSe Superconductor}},\ }\href
  {https://doi.org/10.1103/PhysRevLett.113.237001} {\bibfield  {journal}
  {\bibinfo  {journal} {Phys. Rev. Lett.}\ }\textbf {\bibinfo {volume} {113}},\
  \bibinfo {pages} {237001} (\bibinfo {year} {2014})}\BibitemShut {NoStop}%
\bibitem [{\citenamefont {Shimojima}\ \emph {et~al.}(2014)\citenamefont
  {Shimojima}, \citenamefont {Suzuki}, \citenamefont {Sonobe}, \citenamefont
  {Nakamura}, \citenamefont {Sakano}, \citenamefont {Omachi}, \citenamefont
  {Yoshioka}, \citenamefont {Kuwata-Gonokami}, \citenamefont {Ono},
  \citenamefont {Kumigashira}, \citenamefont {B\"ohmer}, \citenamefont {Hardy},
  \citenamefont {Wolf}, \citenamefont {Meingast}, \citenamefont {L\"ohneysen},
  \citenamefont {Ikeda},\ and\ \citenamefont {Ishizaka}}]{PhysRevB.90.121111}%
  \BibitemOpen
  \bibfield  {author} {\bibinfo {author} {\bibfnamefont {T.}~\bibnamefont
  {Shimojima}}, \bibinfo {author} {\bibfnamefont {Y.}~\bibnamefont {Suzuki}},
  \bibinfo {author} {\bibfnamefont {T.}~\bibnamefont {Sonobe}}, \bibinfo
  {author} {\bibfnamefont {A.}~\bibnamefont {Nakamura}}, \bibinfo {author}
  {\bibfnamefont {M.}~\bibnamefont {Sakano}}, \bibinfo {author} {\bibfnamefont
  {J.}~\bibnamefont {Omachi}}, \bibinfo {author} {\bibfnamefont
  {K.}~\bibnamefont {Yoshioka}}, \bibinfo {author} {\bibfnamefont
  {M.}~\bibnamefont {Kuwata-Gonokami}}, \bibinfo {author} {\bibfnamefont
  {K.}~\bibnamefont {Ono}}, \bibinfo {author} {\bibfnamefont {H.}~\bibnamefont
  {Kumigashira}}, \bibinfo {author} {\bibfnamefont {A.~E.}\ \bibnamefont
  {B\"ohmer}}, \bibinfo {author} {\bibfnamefont {F.}~\bibnamefont {Hardy}},
  \bibinfo {author} {\bibfnamefont {T.}~\bibnamefont {Wolf}}, \bibinfo {author}
  {\bibfnamefont {C.}~\bibnamefont {Meingast}}, \bibinfo {author}
  {\bibfnamefont {H.~v.}\ \bibnamefont {L\"ohneysen}}, \bibinfo {author}
  {\bibfnamefont {H.}~\bibnamefont {Ikeda}},\ and\ \bibinfo {author}
  {\bibfnamefont {K.}~\bibnamefont {Ishizaka}},\ }\bibfield  {title} {\bibinfo
  {title} {{Lifting of $xz/yz$ orbital degeneracy at the structural transition
  in detwinned FeSe}},\ }\href {https://doi.org/10.1103/PhysRevB.90.121111}
  {\bibfield  {journal} {\bibinfo  {journal} {Phys. Rev. B}\ }\textbf {\bibinfo
  {volume} {90}},\ \bibinfo {pages} {121111} (\bibinfo {year}
  {2014})}\BibitemShut {NoStop}%
\bibitem [{\citenamefont {Liu}\ \emph {et~al.}(2018)\citenamefont {Liu},
  \citenamefont {Li}, \citenamefont {Huang}, \citenamefont {Lei}, \citenamefont
  {Wang}, \citenamefont {Wu}, \citenamefont {Shen}, \citenamefont {Gao},
  \citenamefont {Zhang}, \citenamefont {Liu}, \citenamefont {Hu}, \citenamefont
  {Xu}, \citenamefont {Liang}, \citenamefont {Liu}, \citenamefont {Ai},
  \citenamefont {Zhao}, \citenamefont {He}, \citenamefont {Yu}, \citenamefont
  {Liu}, \citenamefont {Mao}, \citenamefont {Dong}, \citenamefont {Jia},
  \citenamefont {Zhang}, \citenamefont {Zhang}, \citenamefont {Yang},
  \citenamefont {Wang}, \citenamefont {Peng}, \citenamefont {Shi},
  \citenamefont {Hu}, \citenamefont {Xiang}, \citenamefont {Chen},
  \citenamefont {Xu}, \citenamefont {Chen},\ and\ \citenamefont
  {Zhou}}]{PhysRevX.8.031033}%
  \BibitemOpen
  \bibfield  {author} {\bibinfo {author} {\bibfnamefont {D.}~\bibnamefont
  {Liu}}, \bibinfo {author} {\bibfnamefont {C.}~\bibnamefont {Li}}, \bibinfo
  {author} {\bibfnamefont {J.}~\bibnamefont {Huang}}, \bibinfo {author}
  {\bibfnamefont {B.}~\bibnamefont {Lei}}, \bibinfo {author} {\bibfnamefont
  {L.}~\bibnamefont {Wang}}, \bibinfo {author} {\bibfnamefont {X.}~\bibnamefont
  {Wu}}, \bibinfo {author} {\bibfnamefont {B.}~\bibnamefont {Shen}}, \bibinfo
  {author} {\bibfnamefont {Q.}~\bibnamefont {Gao}}, \bibinfo {author}
  {\bibfnamefont {Y.}~\bibnamefont {Zhang}}, \bibinfo {author} {\bibfnamefont
  {X.}~\bibnamefont {Liu}}, \bibinfo {author} {\bibfnamefont {Y.}~\bibnamefont
  {Hu}}, \bibinfo {author} {\bibfnamefont {Y.}~\bibnamefont {Xu}}, \bibinfo
  {author} {\bibfnamefont {A.}~\bibnamefont {Liang}}, \bibinfo {author}
  {\bibfnamefont {J.}~\bibnamefont {Liu}}, \bibinfo {author} {\bibfnamefont
  {P.}~\bibnamefont {Ai}}, \bibinfo {author} {\bibfnamefont {L.}~\bibnamefont
  {Zhao}}, \bibinfo {author} {\bibfnamefont {S.}~\bibnamefont {He}}, \bibinfo
  {author} {\bibfnamefont {L.}~\bibnamefont {Yu}}, \bibinfo {author}
  {\bibfnamefont {G.}~\bibnamefont {Liu}}, \bibinfo {author} {\bibfnamefont
  {Y.}~\bibnamefont {Mao}}, \bibinfo {author} {\bibfnamefont {X.}~\bibnamefont
  {Dong}}, \bibinfo {author} {\bibfnamefont {X.}~\bibnamefont {Jia}}, \bibinfo
  {author} {\bibfnamefont {F.}~\bibnamefont {Zhang}}, \bibinfo {author}
  {\bibfnamefont {S.}~\bibnamefont {Zhang}}, \bibinfo {author} {\bibfnamefont
  {F.}~\bibnamefont {Yang}}, \bibinfo {author} {\bibfnamefont {Z.}~\bibnamefont
  {Wang}}, \bibinfo {author} {\bibfnamefont {Q.}~\bibnamefont {Peng}}, \bibinfo
  {author} {\bibfnamefont {Y.}~\bibnamefont {Shi}}, \bibinfo {author}
  {\bibfnamefont {J.}~\bibnamefont {Hu}}, \bibinfo {author} {\bibfnamefont
  {T.}~\bibnamefont {Xiang}}, \bibinfo {author} {\bibfnamefont
  {X.}~\bibnamefont {Chen}}, \bibinfo {author} {\bibfnamefont {Z.}~\bibnamefont
  {Xu}}, \bibinfo {author} {\bibfnamefont {C.}~\bibnamefont {Chen}},\ and\
  \bibinfo {author} {\bibfnamefont {X.~J.}\ \bibnamefont {Zhou}},\ }\bibfield
  {title} {\bibinfo {title} {{Orbital Origin of Extremely Anisotropic
  Superconducting Gap in Nematic Phase of FeSe Superconductor}},\ }\href
  {https://doi.org/10.1103/PhysRevX.8.031033} {\bibfield  {journal} {\bibinfo
  {journal} {Phys. Rev. X}\ }\textbf {\bibinfo {volume} {8}},\ \bibinfo {pages}
  {031033} (\bibinfo {year} {2018})}\BibitemShut {NoStop}%
\bibitem [{\citenamefont {Sprau}\ \emph {et~al.}(2017)\citenamefont {Sprau},
  \citenamefont {Kostin}, \citenamefont {Kreisel}, \citenamefont {Böhmer},
  \citenamefont {Taufour}, \citenamefont {Canfield}, \citenamefont {Mukherjee},
  \citenamefont {Hirschfeld}, \citenamefont {Andersen},\ and\ \citenamefont
  {Davis}}]{doi:10.1126/science.aal1575}%
  \BibitemOpen
  \bibfield  {author} {\bibinfo {author} {\bibfnamefont {P.~O.}\ \bibnamefont
  {Sprau}}, \bibinfo {author} {\bibfnamefont {A.}~\bibnamefont {Kostin}},
  \bibinfo {author} {\bibfnamefont {A.}~\bibnamefont {Kreisel}}, \bibinfo
  {author} {\bibfnamefont {A.~E.}\ \bibnamefont {Böhmer}}, \bibinfo {author}
  {\bibfnamefont {V.}~\bibnamefont {Taufour}}, \bibinfo {author} {\bibfnamefont
  {P.~C.}\ \bibnamefont {Canfield}}, \bibinfo {author} {\bibfnamefont
  {S.}~\bibnamefont {Mukherjee}}, \bibinfo {author} {\bibfnamefont {P.~J.}\
  \bibnamefont {Hirschfeld}}, \bibinfo {author} {\bibfnamefont {B.~M.}\
  \bibnamefont {Andersen}},\ and\ \bibinfo {author} {\bibfnamefont {J.~C.~S.}\
  \bibnamefont {Davis}},\ }\bibfield  {title} {\bibinfo {title} {{Discovery of
  orbital-selective Cooper pairing in FeSe}},\ }\href
  {https://doi.org/10.1126/science.aal1575} {\bibfield  {journal} {\bibinfo
  {journal} {Science}\ }\textbf {\bibinfo {volume} {357}},\ \bibinfo {pages}
  {75} (\bibinfo {year} {2017})}\BibitemShut {NoStop}%
\bibitem [{\citenamefont {Song}\ \emph {et~al.}(2011)\citenamefont {Song},
  \citenamefont {Wang}, \citenamefont {Cheng}, \citenamefont {Jiang},
  \citenamefont {Li}, \citenamefont {Zhang}, \citenamefont {Li}, \citenamefont
  {He}, \citenamefont {Wang}, \citenamefont {Jia}, \citenamefont {Hung},
  \citenamefont {Wu}, \citenamefont {Ma}, \citenamefont {Chen},\ and\
  \citenamefont {Xue}}]{doi:10.1126/science.1202226}%
  \BibitemOpen
  \bibfield  {author} {\bibinfo {author} {\bibfnamefont {C.-L.}\ \bibnamefont
  {Song}}, \bibinfo {author} {\bibfnamefont {Y.-L.}\ \bibnamefont {Wang}},
  \bibinfo {author} {\bibfnamefont {P.}~\bibnamefont {Cheng}}, \bibinfo
  {author} {\bibfnamefont {Y.-P.}\ \bibnamefont {Jiang}}, \bibinfo {author}
  {\bibfnamefont {W.}~\bibnamefont {Li}}, \bibinfo {author} {\bibfnamefont
  {T.}~\bibnamefont {Zhang}}, \bibinfo {author} {\bibfnamefont
  {Z.}~\bibnamefont {Li}}, \bibinfo {author} {\bibfnamefont {K.}~\bibnamefont
  {He}}, \bibinfo {author} {\bibfnamefont {L.}~\bibnamefont {Wang}}, \bibinfo
  {author} {\bibfnamefont {J.-F.}\ \bibnamefont {Jia}}, \bibinfo {author}
  {\bibfnamefont {H.-H.}\ \bibnamefont {Hung}}, \bibinfo {author}
  {\bibfnamefont {C.}~\bibnamefont {Wu}}, \bibinfo {author} {\bibfnamefont
  {X.}~\bibnamefont {Ma}}, \bibinfo {author} {\bibfnamefont {X.}~\bibnamefont
  {Chen}},\ and\ \bibinfo {author} {\bibfnamefont {Q.-K.}\ \bibnamefont
  {Xue}},\ }\bibfield  {title} {\bibinfo {title} {{Direct Observation of Nodes
  and Twofold Symmetry in FeSe Superconductor}},\ }\href
  {https://doi.org/10.1126/science.1202226} {\bibfield  {journal} {\bibinfo
  {journal} {Science}\ }\textbf {\bibinfo {volume} {332}},\ \bibinfo {pages}
  {1410} (\bibinfo {year} {2011})}\BibitemShut {NoStop}%
\bibitem [{\citenamefont {Song}\ \emph {et~al.}(2012)\citenamefont {Song},
  \citenamefont {Wang}, \citenamefont {Jiang}, \citenamefont {Wang},
  \citenamefont {He}, \citenamefont {Chen}, \citenamefont {Hoffman},
  \citenamefont {Ma},\ and\ \citenamefont {Xue}}]{PhysRevLett.109.137004}%
  \BibitemOpen
  \bibfield  {author} {\bibinfo {author} {\bibfnamefont {C.-L.}\ \bibnamefont
  {Song}}, \bibinfo {author} {\bibfnamefont {Y.-L.}\ \bibnamefont {Wang}},
  \bibinfo {author} {\bibfnamefont {Y.-P.}\ \bibnamefont {Jiang}}, \bibinfo
  {author} {\bibfnamefont {L.}~\bibnamefont {Wang}}, \bibinfo {author}
  {\bibfnamefont {K.}~\bibnamefont {He}}, \bibinfo {author} {\bibfnamefont
  {X.}~\bibnamefont {Chen}}, \bibinfo {author} {\bibfnamefont {J.~E.}\
  \bibnamefont {Hoffman}}, \bibinfo {author} {\bibfnamefont {X.-C.}\
  \bibnamefont {Ma}},\ and\ \bibinfo {author} {\bibfnamefont {Q.-K.}\
  \bibnamefont {Xue}},\ }\bibfield  {title} {\bibinfo {title} {{Suppression of
  Superconductivity by Twin Boundaries in FeSe}},\ }\href
  {https://doi.org/10.1103/PhysRevLett.109.137004} {\bibfield  {journal}
  {\bibinfo  {journal} {Phys. Rev. Lett.}\ }\textbf {\bibinfo {volume} {109}},\
  \bibinfo {pages} {137004} (\bibinfo {year} {2012})}\BibitemShut {NoStop}%
\bibitem [{\citenamefont {Hanaguri}\ \emph {et~al.}(2019)\citenamefont
  {Hanaguri}, \citenamefont {Kasahara}, \citenamefont {B\"oker}, \citenamefont
  {Eremin}, \citenamefont {Shibauchi},\ and\ \citenamefont
  {Matsuda}}]{PhysRevLett.122.077001}%
  \BibitemOpen
  \bibfield  {author} {\bibinfo {author} {\bibfnamefont {T.}~\bibnamefont
  {Hanaguri}}, \bibinfo {author} {\bibfnamefont {S.}~\bibnamefont {Kasahara}},
  \bibinfo {author} {\bibfnamefont {J.}~\bibnamefont {B\"oker}}, \bibinfo
  {author} {\bibfnamefont {I.}~\bibnamefont {Eremin}}, \bibinfo {author}
  {\bibfnamefont {T.}~\bibnamefont {Shibauchi}},\ and\ \bibinfo {author}
  {\bibfnamefont {Y.}~\bibnamefont {Matsuda}},\ }\bibfield  {title} {\bibinfo
  {title} {{Quantum Vortex Core and Missing Pseudogap in the Multiband BCS-BEC
  Crossover Superconductor FeSe}},\ }\href
  {https://doi.org/10.1103/PhysRevLett.122.077001} {\bibfield  {journal}
  {\bibinfo  {journal} {Phys. Rev. Lett.}\ }\textbf {\bibinfo {volume} {122}},\
  \bibinfo {pages} {077001} (\bibinfo {year} {2019})}\BibitemShut {NoStop}%
\bibitem [{\citenamefont {Hashimoto}\ \emph {et~al.}(2018)\citenamefont
  {Hashimoto}, \citenamefont {Ota}, \citenamefont {Yamamoto}, \citenamefont
  {Suzuki}, \citenamefont {Shimojima}, \citenamefont {Watanabe}, \citenamefont
  {Chen}, \citenamefont {Kasahara}, \citenamefont {Matsuda}, \citenamefont
  {Shibauchi}, \citenamefont {Okazaki},\ and\ \citenamefont
  {Shin}}]{NatComm.9-282}%
  \BibitemOpen
  \bibfield  {author} {\bibinfo {author} {\bibfnamefont {T.}~\bibnamefont
  {Hashimoto}}, \bibinfo {author} {\bibfnamefont {Y.}~\bibnamefont {Ota}},
  \bibinfo {author} {\bibfnamefont {H.~Q.}\ \bibnamefont {Yamamoto}}, \bibinfo
  {author} {\bibfnamefont {Y.}~\bibnamefont {Suzuki}}, \bibinfo {author}
  {\bibfnamefont {T.}~\bibnamefont {Shimojima}}, \bibinfo {author}
  {\bibfnamefont {S.}~\bibnamefont {Watanabe}}, \bibinfo {author}
  {\bibfnamefont {C.}~\bibnamefont {Chen}}, \bibinfo {author} {\bibfnamefont
  {S.}~\bibnamefont {Kasahara}}, \bibinfo {author} {\bibfnamefont
  {Y.}~\bibnamefont {Matsuda}}, \bibinfo {author} {\bibfnamefont
  {T.}~\bibnamefont {Shibauchi}}, \bibinfo {author} {\bibfnamefont
  {K.}~\bibnamefont {Okazaki}},\ and\ \bibinfo {author} {\bibfnamefont
  {S.}~\bibnamefont {Shin}},\ }\bibfield  {title} {\bibinfo {title}
  {{Superconducting gap anisotropy sensitive to nematic domains in FeSe}},\
  }\href {https://doi.org/10.1038/s41467-017-02739-y} {\bibfield  {journal}
  {\bibinfo  {journal} {Nature Communications}\ }\textbf {\bibinfo {volume}
  {9}},\ \bibinfo {pages} {282} (\bibinfo {year} {2018})}\BibitemShut {NoStop}%
\bibitem [{\citenamefont {Song}\ \emph {et~al.}(2023)\citenamefont {Song},
  \citenamefont {Hua}, \citenamefont {Bell}, \citenamefont {Ko}, \citenamefont
  {Fangohr}, \citenamefont {Yan}, \citenamefont {Halász}, \citenamefont
  {Dumitrescu}, \citenamefont {Lawrie},\ and\ \citenamefont
  {Maksymovych}}]{doi:10.1021/acs.nanolett.3c00125}%
  \BibitemOpen
  \bibfield  {author} {\bibinfo {author} {\bibfnamefont {S.~Y.}\ \bibnamefont
  {Song}}, \bibinfo {author} {\bibfnamefont {C.}~\bibnamefont {Hua}}, \bibinfo
  {author} {\bibfnamefont {L.}~\bibnamefont {Bell}}, \bibinfo {author}
  {\bibfnamefont {W.}~\bibnamefont {Ko}}, \bibinfo {author} {\bibfnamefont
  {H.}~\bibnamefont {Fangohr}}, \bibinfo {author} {\bibfnamefont
  {J.}~\bibnamefont {Yan}}, \bibinfo {author} {\bibfnamefont {G.~B.}\
  \bibnamefont {Halász}}, \bibinfo {author} {\bibfnamefont {E.~F.}\
  \bibnamefont {Dumitrescu}}, \bibinfo {author} {\bibfnamefont {B.~J.}\
  \bibnamefont {Lawrie}},\ and\ \bibinfo {author} {\bibfnamefont
  {P.}~\bibnamefont {Maksymovych}},\ }\bibfield  {title} {\bibinfo {title}
  {{Nematically Templated Vortex Lattices in Superconducting FeSe}},\
  }\href@noop {} {\bibfield  {journal} {\bibinfo  {journal} {Nano Letters}\
  }\textbf {\bibinfo {volume} {23}},\ \bibinfo {pages} {2822} (\bibinfo {year}
  {2023})}\BibitemShut {NoStop}%
\bibitem [{\citenamefont {Putilov}\ \emph {et~al.}(2019)\citenamefont
  {Putilov}, \citenamefont {Di~Giorgio}, \citenamefont {Vadimov}, \citenamefont
  {Trainer}, \citenamefont {Lechner}, \citenamefont {Curtis}, \citenamefont
  {Abdel-Hafiez}, \citenamefont {Volkova}, \citenamefont {Vasiliev},
  \citenamefont {Chareev}, \citenamefont {Karapetrov}, \citenamefont
  {Koshelev}, \citenamefont {Aladyshkin}, \citenamefont {Mel'nikov},\ and\
  \citenamefont {Iavarone}}]{PhysRevB.99.144514}%
  \BibitemOpen
  \bibfield  {author} {\bibinfo {author} {\bibfnamefont {A.~V.}\ \bibnamefont
  {Putilov}}, \bibinfo {author} {\bibfnamefont {C.}~\bibnamefont {Di~Giorgio}},
  \bibinfo {author} {\bibfnamefont {V.~L.}\ \bibnamefont {Vadimov}}, \bibinfo
  {author} {\bibfnamefont {D.~J.}\ \bibnamefont {Trainer}}, \bibinfo {author}
  {\bibfnamefont {E.~M.}\ \bibnamefont {Lechner}}, \bibinfo {author}
  {\bibfnamefont {J.~L.}\ \bibnamefont {Curtis}}, \bibinfo {author}
  {\bibfnamefont {M.}~\bibnamefont {Abdel-Hafiez}}, \bibinfo {author}
  {\bibfnamefont {O.~S.}\ \bibnamefont {Volkova}}, \bibinfo {author}
  {\bibfnamefont {A.~N.}\ \bibnamefont {Vasiliev}}, \bibinfo {author}
  {\bibfnamefont {D.~A.}\ \bibnamefont {Chareev}}, \bibinfo {author}
  {\bibfnamefont {G.}~\bibnamefont {Karapetrov}}, \bibinfo {author}
  {\bibfnamefont {A.~E.}\ \bibnamefont {Koshelev}}, \bibinfo {author}
  {\bibfnamefont {A.~Y.}\ \bibnamefont {Aladyshkin}}, \bibinfo {author}
  {\bibfnamefont {A.~S.}\ \bibnamefont {Mel'nikov}},\ and\ \bibinfo {author}
  {\bibfnamefont {M.}~\bibnamefont {Iavarone}},\ }\bibfield  {title} {\bibinfo
  {title} {{Vortex-core properties and vortex-lattice transformation in
  FeSe}},\ }\href {https://doi.org/10.1103/PhysRevB.99.144514} {\bibfield
  {journal} {\bibinfo  {journal} {Phys. Rev. B}\ }\textbf {\bibinfo {volume}
  {99}},\ \bibinfo {pages} {144514} (\bibinfo {year} {2019})}\BibitemShut
  {NoStop}%
\bibitem [{\citenamefont {Watashige}\ \emph {et~al.}(2015)\citenamefont
  {Watashige}, \citenamefont {Tsutsumi}, \citenamefont {Hanaguri},
  \citenamefont {Kohsaka}, \citenamefont {Kasahara}, \citenamefont {Furusaki},
  \citenamefont {Sigrist}, \citenamefont {Meingast}, \citenamefont {Wolf},
  \citenamefont {L\"ohneysen}, \citenamefont {Shibauchi},\ and\ \citenamefont
  {Matsuda}}]{PhysRevX.5.031022}%
  \BibitemOpen
  \bibfield  {author} {\bibinfo {author} {\bibfnamefont {T.}~\bibnamefont
  {Watashige}}, \bibinfo {author} {\bibfnamefont {Y.}~\bibnamefont {Tsutsumi}},
  \bibinfo {author} {\bibfnamefont {T.}~\bibnamefont {Hanaguri}}, \bibinfo
  {author} {\bibfnamefont {Y.}~\bibnamefont {Kohsaka}}, \bibinfo {author}
  {\bibfnamefont {S.}~\bibnamefont {Kasahara}}, \bibinfo {author}
  {\bibfnamefont {A.}~\bibnamefont {Furusaki}}, \bibinfo {author}
  {\bibfnamefont {M.}~\bibnamefont {Sigrist}}, \bibinfo {author} {\bibfnamefont
  {C.}~\bibnamefont {Meingast}}, \bibinfo {author} {\bibfnamefont
  {T.}~\bibnamefont {Wolf}}, \bibinfo {author} {\bibfnamefont {H.~v.}\
  \bibnamefont {L\"ohneysen}}, \bibinfo {author} {\bibfnamefont
  {T.}~\bibnamefont {Shibauchi}},\ and\ \bibinfo {author} {\bibfnamefont
  {Y.}~\bibnamefont {Matsuda}},\ }\bibfield  {title} {\bibinfo {title}
  {{Evidence for Time-Reversal Symmetry Breaking of the Superconducting State
  near Twin-Boundary Interfaces in FeSe Revealed by Scanning Tunneling
  Spectroscopy}},\ }\href {https://doi.org/10.1103/PhysRevX.5.031022}
  {\bibfield  {journal} {\bibinfo  {journal} {Phys. Rev. X}\ }\textbf {\bibinfo
  {volume} {5}},\ \bibinfo {pages} {031022} (\bibinfo {year}
  {2015})}\BibitemShut {NoStop}%
\bibitem [{\citenamefont {Sigrist}\ \emph {et~al.}(1996)\citenamefont
  {Sigrist}, \citenamefont {Kuboki}, \citenamefont {Lee}, \citenamefont
  {Millis},\ and\ \citenamefont {Rice}}]{PhysRevB.53.2835}%
  \BibitemOpen
  \bibfield  {author} {\bibinfo {author} {\bibfnamefont {M.}~\bibnamefont
  {Sigrist}}, \bibinfo {author} {\bibfnamefont {K.}~\bibnamefont {Kuboki}},
  \bibinfo {author} {\bibfnamefont {P.~A.}\ \bibnamefont {Lee}}, \bibinfo
  {author} {\bibfnamefont {A.~J.}\ \bibnamefont {Millis}},\ and\ \bibinfo
  {author} {\bibfnamefont {T.~M.}\ \bibnamefont {Rice}},\ }\bibfield  {title}
  {\bibinfo {title} {{Influence of twin boundaries on Josephson junctions
  between high-temperature and conventional superconductors}},\ }\href
  {https://doi.org/10.1103/PhysRevB.53.2835} {\bibfield  {journal} {\bibinfo
  {journal} {Phys. Rev. B}\ }\textbf {\bibinfo {volume} {53}},\ \bibinfo
  {pages} {2835} (\bibinfo {year} {1996})}\BibitemShut {NoStop}%
\bibitem [{\citenamefont {Lu}\ \emph {et~al.}(2018)\citenamefont {Lu},
  \citenamefont {Lv}, \citenamefont {Li}, \citenamefont {Zhu}, \citenamefont
  {Wang}, \citenamefont {Wang},\ and\ \citenamefont {Wu}}]{npjQM.3-12}%
  \BibitemOpen
  \bibfield  {author} {\bibinfo {author} {\bibfnamefont {D.-C.}\ \bibnamefont
  {Lu}}, \bibinfo {author} {\bibfnamefont {Y.-Y.}\ \bibnamefont {Lv}}, \bibinfo
  {author} {\bibfnamefont {J.}~\bibnamefont {Li}}, \bibinfo {author}
  {\bibfnamefont {B.-Y.}\ \bibnamefont {Zhu}}, \bibinfo {author} {\bibfnamefont
  {Q.-H.}\ \bibnamefont {Wang}}, \bibinfo {author} {\bibfnamefont {H.-B.}\
  \bibnamefont {Wang}},\ and\ \bibinfo {author} {\bibfnamefont {P.-H.}\
  \bibnamefont {Wu}},\ }\bibfield  {title} {\bibinfo {title} {{Elliptical
  vortex and oblique vortex lattice in the FeSe superconductor based on the
  nematicity and mixed superconducting orders}},\ }\href
  {https://doi.org/10.1038/s41535-018-0087-2} {\bibfield  {journal} {\bibinfo
  {journal} {npj Quantum Materials}\ }\textbf {\bibinfo {volume} {3}},\
  \bibinfo {pages} {12} (\bibinfo {year} {2018})}\BibitemShut {NoStop}%
\bibitem [{\citenamefont {Noda}\ \emph {et~al.}(2025)\citenamefont {Noda},
  \citenamefont {Adachi},\ and\ \citenamefont
  {Ichioka}}]{doi:10.7566/JPSJ.94.023702}%
  \BibitemOpen
  \bibfield  {author} {\bibinfo {author} {\bibfnamefont {S.}~\bibnamefont
  {Noda}}, \bibinfo {author} {\bibfnamefont {H.}~\bibnamefont {Adachi}},\ and\
  \bibinfo {author} {\bibfnamefont {M.}~\bibnamefont {Ichioka}},\ }\bibfield
  {title} {\bibinfo {title} {Fractional vortex array realized at twin boundary
  in a nematic superconductor},\ }\href
  {https://doi.org/10.7566/JPSJ.94.023702} {\bibfield  {journal} {\bibinfo
  {journal} {Journal of the Physical Society of Japan}\ }\textbf {\bibinfo
  {volume} {94}},\ \bibinfo {pages} {023702} (\bibinfo {year}
  {2025})}\BibitemShut {NoStop}%
\bibitem [{\citenamefont {Severino}\ \emph {et~al.}(2024)\citenamefont
  {Severino}, \citenamefont {Mininni}, \citenamefont {Fradkin}, \citenamefont
  {Bekeris}, \citenamefont {Pasquini},\ and\ \citenamefont
  {Lozano}}]{PhysRevB.109.094513}%
  \BibitemOpen
  \bibfield  {author} {\bibinfo {author} {\bibfnamefont {R.~S.}\ \bibnamefont
  {Severino}}, \bibinfo {author} {\bibfnamefont {P.~D.}\ \bibnamefont
  {Mininni}}, \bibinfo {author} {\bibfnamefont {E.}~\bibnamefont {Fradkin}},
  \bibinfo {author} {\bibfnamefont {V.}~\bibnamefont {Bekeris}}, \bibinfo
  {author} {\bibfnamefont {G.}~\bibnamefont {Pasquini}},\ and\ \bibinfo
  {author} {\bibfnamefont {G.~S.}\ \bibnamefont {Lozano}},\ }\bibfield  {title}
  {\bibinfo {title} {{Ginzburg-Landau approach to the vortex--domain wall
  interaction in superconductors with nematic order}},\ }\href
  {https://doi.org/10.1103/PhysRevB.109.094513} {\bibfield  {journal} {\bibinfo
   {journal} {Phys. Rev. B}\ }\textbf {\bibinfo {volume} {109}},\ \bibinfo
  {pages} {094513} (\bibinfo {year} {2024})}\BibitemShut {NoStop}%
\bibitem [{\citenamefont {Sigrist}\ and\ \citenamefont
  {Agterberg}(1999)}]{10.1143/PTP.102.965}%
  \BibitemOpen
  \bibfield  {author} {\bibinfo {author} {\bibfnamefont {M.}~\bibnamefont
  {Sigrist}}\ and\ \bibinfo {author} {\bibfnamefont {D.~F.}\ \bibnamefont
  {Agterberg}},\ }\bibfield  {title} {\bibinfo {title} {{The Role of Domain
  Walls on the Vortex Creep Dynamics in Unconventional Superconductors}},\
  }\href {https://doi.org/10.1143/PTP.102.965} {\bibfield  {journal} {\bibinfo
  {journal} {Progress of Theoretical Physics}\ }\textbf {\bibinfo {volume}
  {102}},\ \bibinfo {pages} {965} (\bibinfo {year} {1999})}\BibitemShut
  {NoStop}%
\bibitem [{\citenamefont {Talkachov}\ \emph {et~al.}(2025)\citenamefont
  {Talkachov}, \citenamefont {Kaushik},\ and\ \citenamefont
  {Babaev}}]{PhysRevB.111.054513}%
  \BibitemOpen
  \bibfield  {author} {\bibinfo {author} {\bibfnamefont {A.}~\bibnamefont
  {Talkachov}}, \bibinfo {author} {\bibfnamefont {S.}~\bibnamefont {Kaushik}},\
  and\ \bibinfo {author} {\bibfnamefont {E.}~\bibnamefont {Babaev}},\
  }\bibfield  {title} {\bibinfo {title} {Microscopic approach to the problem of
  enhancement and suppression of superconductivity on twinning planes},\ }\href
  {https://doi.org/10.1103/PhysRevB.111.054513} {\bibfield  {journal} {\bibinfo
   {journal} {Phys. Rev. B}\ }\textbf {\bibinfo {volume} {111}},\ \bibinfo
  {pages} {054513} (\bibinfo {year} {2025})}\BibitemShut {NoStop}%
\bibitem [{\citenamefont {Feder}\ \emph {et~al.}(1997)\citenamefont {Feder},
  \citenamefont {Beardsall}, \citenamefont {Berlinsky},\ and\ \citenamefont
  {Kallin}}]{PhysRevB.56.R5751}%
  \BibitemOpen
  \bibfield  {author} {\bibinfo {author} {\bibfnamefont {D.~L.}\ \bibnamefont
  {Feder}}, \bibinfo {author} {\bibfnamefont {A.}~\bibnamefont {Beardsall}},
  \bibinfo {author} {\bibfnamefont {A.~J.}\ \bibnamefont {Berlinsky}},\ and\
  \bibinfo {author} {\bibfnamefont {C.}~\bibnamefont {Kallin}},\ }\bibfield
  {title} {\bibinfo {title} {{Twin boundaries in $d$-wave superconductors}},\
  }\href {https://doi.org/10.1103/PhysRevB.56.R5751} {\bibfield  {journal}
  {\bibinfo  {journal} {Phys. Rev. B}\ }\textbf {\bibinfo {volume} {56}},\
  \bibinfo {pages} {R5751} (\bibinfo {year} {1997})}\BibitemShut {NoStop}%
\bibitem [{\citenamefont {Hung}\ \emph {et~al.}(2012)\citenamefont {Hung},
  \citenamefont {Song}, \citenamefont {Chen}, \citenamefont {Ma}, \citenamefont
  {Xue},\ and\ \citenamefont {Wu}}]{PhysRevB.85.104510}%
  \BibitemOpen
  \bibfield  {author} {\bibinfo {author} {\bibfnamefont {H.-H.}\ \bibnamefont
  {Hung}}, \bibinfo {author} {\bibfnamefont {C.-L.}\ \bibnamefont {Song}},
  \bibinfo {author} {\bibfnamefont {X.}~\bibnamefont {Chen}}, \bibinfo {author}
  {\bibfnamefont {X.}~\bibnamefont {Ma}}, \bibinfo {author} {\bibfnamefont
  {Q.-k.}\ \bibnamefont {Xue}},\ and\ \bibinfo {author} {\bibfnamefont
  {C.}~\bibnamefont {Wu}},\ }\bibfield  {title} {\bibinfo {title} {{Anisotropic
  vortex lattice structures in the FeSe superconductor}},\ }\href
  {https://doi.org/10.1103/PhysRevB.85.104510} {\bibfield  {journal} {\bibinfo
  {journal} {Phys. Rev. B}\ }\textbf {\bibinfo {volume} {85}},\ \bibinfo
  {pages} {104510} (\bibinfo {year} {2012})}\BibitemShut {NoStop}%
\bibitem [{\citenamefont {Ichioka}\ and\ \citenamefont
  {Adachi}(2026)}]{27tp-xzjb}%
  \BibitemOpen
  \bibfield  {author} {\bibinfo {author} {\bibfnamefont {M.}~\bibnamefont
  {Ichioka}}\ and\ \bibinfo {author} {\bibfnamefont {H.}~\bibnamefont
  {Adachi}},\ }\bibfield  {title} {\bibinfo {title} {{Twofold symmetric vortex
  core states in nematic superconductors with anisotropic Fermi surface and
  pairing functions}},\ }\href {https://doi.org/10.1103/27tp-xzjb} {\bibfield
  {journal} {\bibinfo  {journal} {Phys. Rev. B}\ }\textbf {\bibinfo {volume}
  {113}},\ \bibinfo {pages} {094520} (\bibinfo {year} {2026})}\BibitemShut
  {NoStop}%
\bibitem [{\citenamefont {Sera}\ \emph {et~al.}(2020)\citenamefont {Sera},
  \citenamefont {Ueda}, \citenamefont {Adachi},\ and\ \citenamefont
  {Ichioka}}]{sym12010175}%
  \BibitemOpen
  \bibfield  {author} {\bibinfo {author} {\bibfnamefont {Y.}~\bibnamefont
  {Sera}}, \bibinfo {author} {\bibfnamefont {T.}~\bibnamefont {Ueda}}, \bibinfo
  {author} {\bibfnamefont {H.}~\bibnamefont {Adachi}},\ and\ \bibinfo {author}
  {\bibfnamefont {M.}~\bibnamefont {Ichioka}},\ }\bibfield  {title} {\bibinfo
  {title} {Relation of superconducting pairing symmetry and non-magnetic
  impurity effects in vortex states},\ }\href@noop {} {\bibfield  {journal}
  {\bibinfo  {journal} {Symmetry}\ }\textbf {\bibinfo {volume} {12}},\ \bibinfo
  {pages} {175} (\bibinfo {year} {2020})}\BibitemShut {NoStop}%
\bibitem [{\citenamefont {Wang}\ and\ \citenamefont
  {MacDonald}(1995)}]{PhysRevB.52.R3876}%
  \BibitemOpen
  \bibfield  {author} {\bibinfo {author} {\bibfnamefont {Y.}~\bibnamefont
  {Wang}}\ and\ \bibinfo {author} {\bibfnamefont {A.~H.}\ \bibnamefont
  {MacDonald}},\ }\bibfield  {title} {\bibinfo {title} {{Mixed-state
  quasiparticle spectrum for $d$-wave superconductors}},\ }\href
  {https://doi.org/10.1103/PhysRevB.52.R3876} {\bibfield  {journal} {\bibinfo
  {journal} {Phys. Rev. B}\ }\textbf {\bibinfo {volume} {52}},\ \bibinfo
  {pages} {R3876} (\bibinfo {year} {1995})}\BibitemShut {NoStop}%
\bibitem [{\citenamefont {Takigawa}\ \emph {et~al.}(2004)\citenamefont
  {Takigawa}, \citenamefont {Ichioka},\ and\ \citenamefont
  {Machida}}]{doi:10.1143/JPSJ.73.450}%
  \BibitemOpen
  \bibfield  {author} {\bibinfo {author} {\bibfnamefont {M.}~\bibnamefont
  {Takigawa}}, \bibinfo {author} {\bibfnamefont {M.}~\bibnamefont {Ichioka}},\
  and\ \bibinfo {author} {\bibfnamefont {K.}~\bibnamefont {Machida}},\
  }\bibfield  {title} {\bibinfo {title} {Quasiparticle structure in
  antiferromagnetism around the vortex and nuclear magnetic relaxation time},\
  }\href {https://doi.org/10.1143/JPSJ.73.450} {\bibfield  {journal} {\bibinfo
  {journal} {Journal of the Physical Society of Japan}\ }\textbf {\bibinfo
  {volume} {73}},\ \bibinfo {pages} {450} (\bibinfo {year} {2004})}\BibitemShut
  {NoStop}%
\bibitem [{\citenamefont {Hayashi}\ \emph {et~al.}(1998)\citenamefont
  {Hayashi}, \citenamefont {Isoshima}, \citenamefont {Ichioka},\ and\
  \citenamefont {Machida}}]{PhysRevLett.80.2921}%
  \BibitemOpen
  \bibfield  {author} {\bibinfo {author} {\bibfnamefont {N.}~\bibnamefont
  {Hayashi}}, \bibinfo {author} {\bibfnamefont {T.}~\bibnamefont {Isoshima}},
  \bibinfo {author} {\bibfnamefont {M.}~\bibnamefont {Ichioka}},\ and\ \bibinfo
  {author} {\bibfnamefont {K.}~\bibnamefont {Machida}},\ }\bibfield  {title}
  {\bibinfo {title} {Low-lying quasiparticle excitations around a vortex core
  in quantum limit},\ }\href {https://doi.org/10.1103/PhysRevLett.80.2921}
  {\bibfield  {journal} {\bibinfo  {journal} {Phys. Rev. Lett.}\ }\textbf
  {\bibinfo {volume} {80}},\ \bibinfo {pages} {2921} (\bibinfo {year}
  {1998})}\BibitemShut {NoStop}%
\bibitem [{\citenamefont {Kaneko}\ \emph {et~al.}(2012)\citenamefont {Kaneko},
  \citenamefont {Matsuba}, \citenamefont {Hafiz}, \citenamefont {Yamasaki},
  \citenamefont {Kakizaki}, \citenamefont {Nishida}, \citenamefont {Takeya},
  \citenamefont {Hirata}, \citenamefont {Kawakami}, \citenamefont {Mizushima},\
  and\ \citenamefont {Machida}}]{Kaneko}%
  \BibitemOpen
  \bibfield  {author} {\bibinfo {author} {\bibfnamefont {S.-i.}\ \bibnamefont
  {Kaneko}}, \bibinfo {author} {\bibfnamefont {K.}~\bibnamefont {Matsuba}},
  \bibinfo {author} {\bibfnamefont {M.}~\bibnamefont {Hafiz}}, \bibinfo
  {author} {\bibfnamefont {K.}~\bibnamefont {Yamasaki}}, \bibinfo {author}
  {\bibfnamefont {E.}~\bibnamefont {Kakizaki}}, \bibinfo {author}
  {\bibfnamefont {N.}~\bibnamefont {Nishida}}, \bibinfo {author} {\bibfnamefont
  {H.}~\bibnamefont {Takeya}}, \bibinfo {author} {\bibfnamefont
  {K.}~\bibnamefont {Hirata}}, \bibinfo {author} {\bibfnamefont
  {T.}~\bibnamefont {Kawakami}}, \bibinfo {author} {\bibfnamefont
  {T.}~\bibnamefont {Mizushima}},\ and\ \bibinfo {author} {\bibfnamefont
  {K.}~\bibnamefont {Machida}},\ }\bibfield  {title} {\bibinfo {title}
  {{Quantum Limiting Behaviors of a Vortex Core in an Anisotropic Gap
  Superconductor}},\ }\href {https://doi.org/10.1143/JPSJ.81.063701} {\bibfield
   {journal} {\bibinfo  {journal} {Journal of the Physical Society of Japan}\
  }\textbf {\bibinfo {volume} {81}},\ \bibinfo {pages} {063701} (\bibinfo
  {year} {2012})}\BibitemShut {NoStop}%
\bibitem [{\citenamefont {Caroli}\ \emph {et~al.}(1964)\citenamefont {Caroli},
  \citenamefont {{P. G. {De Gennes}}},\ and\ \citenamefont
  {Matricon}}]{CAROLI1964307}%
  \BibitemOpen
  \bibfield  {author} {\bibinfo {author} {\bibfnamefont {C.}~\bibnamefont
  {Caroli}}, \bibinfo {author} {\bibnamefont {{P. G. {De Gennes}}}},\ and\
  \bibinfo {author} {\bibfnamefont {J.}~\bibnamefont {Matricon}},\ }\bibfield
  {title} {\bibinfo {title} {{Bound Fermion states on a vortex line in a type
  II superconductor}},\ }\href
  {https://doi.org/https://doi.org/10.1016/0031-9163(64)90375-0} {\bibfield
  {journal} {\bibinfo  {journal} {Physics Letters}\ }\textbf {\bibinfo {volume}
  {9}},\ \bibinfo {pages} {307} (\bibinfo {year} {1964})}\BibitemShut {NoStop}%
\bibitem [{\citenamefont {Gygi}\ and\ \citenamefont
  {Schl\"uter}(1991)}]{PhysRevB.43.7609}%
  \BibitemOpen
  \bibfield  {author} {\bibinfo {author} {\bibfnamefont {F.}~\bibnamefont
  {Gygi}}\ and\ \bibinfo {author} {\bibfnamefont {M.}~\bibnamefont
  {Schl\"uter}},\ }\bibfield  {title} {\bibinfo {title} {{Self-consistent
  electronic structure of a vortex line in a type-II superconductor}},\ }\href
  {https://doi.org/10.1103/PhysRevB.43.7609} {\bibfield  {journal} {\bibinfo
  {journal} {Phys. Rev. B}\ }\textbf {\bibinfo {volume} {43}},\ \bibinfo
  {pages} {7609} (\bibinfo {year} {1991})}\BibitemShut {NoStop}%
\bibitem [{\citenamefont {Ichioka}\ \emph {et~al.}(1996)\citenamefont
  {Ichioka}, \citenamefont {Hayashi}, \citenamefont {Enomoto},\ and\
  \citenamefont {Machida}}]{PhysRevB.53.15316}%
  \BibitemOpen
  \bibfield  {author} {\bibinfo {author} {\bibfnamefont {M.}~\bibnamefont
  {Ichioka}}, \bibinfo {author} {\bibfnamefont {N.}~\bibnamefont {Hayashi}},
  \bibinfo {author} {\bibfnamefont {N.}~\bibnamefont {Enomoto}},\ and\ \bibinfo
  {author} {\bibfnamefont {K.}~\bibnamefont {Machida}},\ }\bibfield  {title}
  {\bibinfo {title} {{Vortex structure in $d$-wave superconductors}},\ }\href
  {https://doi.org/10.1103/PhysRevB.53.15316} {\bibfield  {journal} {\bibinfo
  {journal} {Phys. Rev. B}\ }\textbf {\bibinfo {volume} {53}},\ \bibinfo
  {pages} {15316} (\bibinfo {year} {1996})}\BibitemShut {NoStop}%
\bibitem [{\citenamefont {Ichioka}\ \emph {et~al.}(1999)\citenamefont
  {Ichioka}, \citenamefont {Hasegawa},\ and\ \citenamefont
  {Machida}}]{PhysRevB.59.8902}%
  \BibitemOpen
  \bibfield  {author} {\bibinfo {author} {\bibfnamefont {M.}~\bibnamefont
  {Ichioka}}, \bibinfo {author} {\bibfnamefont {A.}~\bibnamefont {Hasegawa}},\
  and\ \bibinfo {author} {\bibfnamefont {K.}~\bibnamefont {Machida}},\
  }\bibfield  {title} {\bibinfo {title} {{Field dependence of the vortex
  structure in $d$-wave and $s$-wave superconductors}},\ }\href
  {https://doi.org/10.1103/PhysRevB.59.8902} {\bibfield  {journal} {\bibinfo
  {journal} {Phys. Rev. B}\ }\textbf {\bibinfo {volume} {59}},\ \bibinfo
  {pages} {8902} (\bibinfo {year} {1999})}\BibitemShut {NoStop}%
\end{thebibliography}

%
%


\end{document}